%% file: main_v1_0_acs.tex
\documentclass[journal=jcisd8,manuscript=article]{achemso}

\usepackage[T1]{fontenc}
\usepackage{lmodern}

\usepackage[version=3]{mhchem} 

\usepackage[ruled,vlined,linesnumbered]{algorithm2e}
\usepackage{amsmath}
\usepackage{amssymb}
\usepackage{booktabs}
\usepackage{enumitem}
\usepackage{graphicx}
\usepackage{listings}
\usepackage{mathtools}
\usepackage{multirow}
\usepackage{placeins}
\usepackage{rotating}
\usepackage{siunitx}
\usepackage{threeparttable}
\usepackage{wrapfig}
\usepackage{xr}
\usepackage{hyperref}
\usepackage{xcolor}

\definecolor{cream}{RGB}{222,217,201}

\newcommand{\argmax}{\mathop{\arg\max}\limits}

\author{David E. Graff}
\affiliation{Department of Chemistry and Chemical Biology, Harvard University, Cambridge, MA 02138}
\alsoaffiliation{Department of Chemical Engineering, MIT, Cambridge, MA 02142}

\author{Matteo Aldeghi}
\affiliation{Department of Chemical Engineering, MIT, Cambridge, MA 02142}

\author{Joseph A. Morrone}
\affiliation{Computational Biology Center, IBM Thomas J. Watson Research Center, Yorktown Heights, NY 10594}
\author{Kirk E. Jordan}
\affiliation{IBM Thomas J. Watson Research Center, Cambridge, Massachusetts 02142}
\author{Edward O. Pyzer-Knapp}
\affiliation{IBM Research Europe, Daresbury, U.K.}

\author{Connor W. Coley}
\email{ccoley@mit.edu}
\affiliation{Department of Chemical Engineering, MIT, Cambridge, MA 02142}
\alsoaffiliation{Department of Electrical Engineering and Computer Science, MIT, Cambridge, MA 02142}

\title[Design Space Pruning]{Self-focusing virtual screening with active design space pruning}

\abbreviations{BO,VS,AL,ML}

\begin{document}

\begin{tocentry}
    \includegraphics[width=8cm]{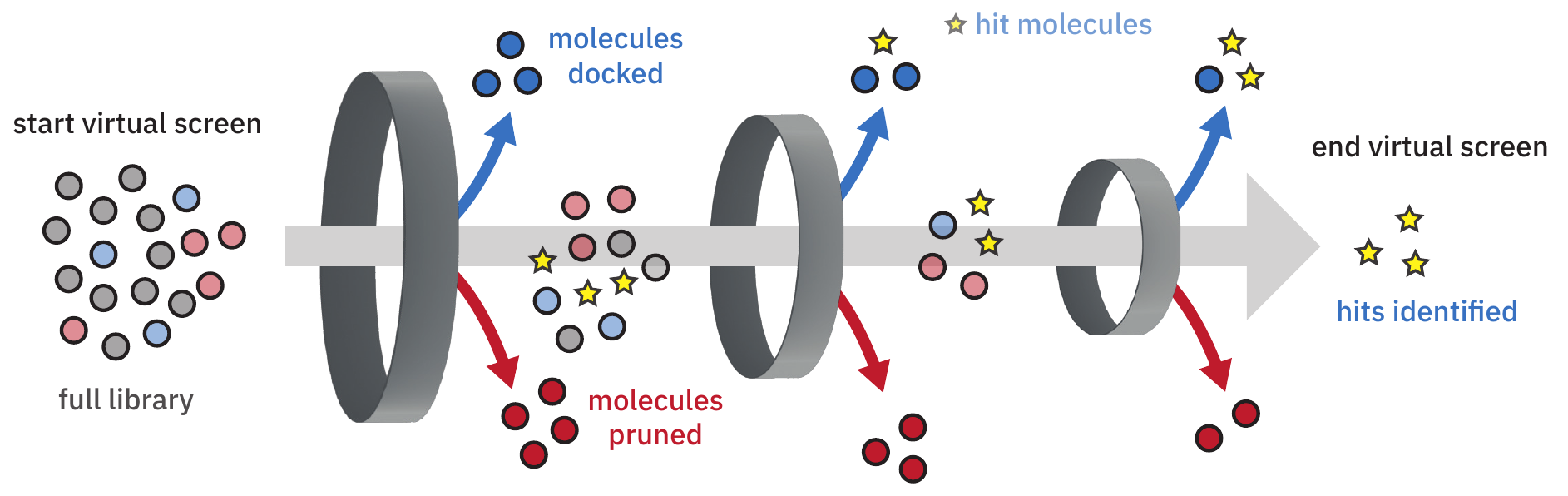}
\end{tocentry}

\begin{abstract}
    High-throughput virtual screening is an indispensable technique utilized in the discovery of small molecules. In cases where the library of molecules is exceedingly large, the cost of an exhaustive virtual screen may be prohibitive. Model-guided optimization has been employed to lower these costs through dramatic increases in sample efficiency compared to random selection. However, these techniques introduce new costs to the workflow through the surrogate model training and inference steps. In this study, we propose an extension to the framework of model-guided optimization that mitigates inferences costs using a technique we refer to as design space pruning (DSP), which irreversibly removes poor-performing candidates from consideration. We study the application of DSP to a variety of optimization tasks and observe significant reductions in overhead costs while exhibiting similar performance to the baseline optimization. DSP represents an attractive extension of model-guided optimization that can limit overhead costs in optimization settings where these costs are non-negligible relative to objective costs, such as docking.
\end{abstract}

\section{Introduction}
The discovery of novel molecules and materials is a central problem that cuts across many domains of science. Broadly, these problems entail the search through a space of possible candidates (``design space'') for those that achieve or optimize a specified set of properties. Depending on the context, the search may involve identifying these candidates from a discrete, enumerated library. High-throughput virtual screening (HTVS) is often employed in these scenarios,\cite{pyzer-knapp_what_2015} but the cost of the technique is directly proportional to both the cost of evaluating a single candidate and the total number of candidates. There is a significant amount of recent work on HTVS for structure-based drug design (SBDD) using computational docking to screen ultra-large chemical libraries exceeding one hundred million or even one billion compounds.\cite{lyu_ultra-large_2019,gorgulla_open-source_2020,gentile_automated_2021,luttens_ultralarge_2022}

\begin{figure}[t!]
    \centering
    \includegraphics[width=0.7\textwidth]{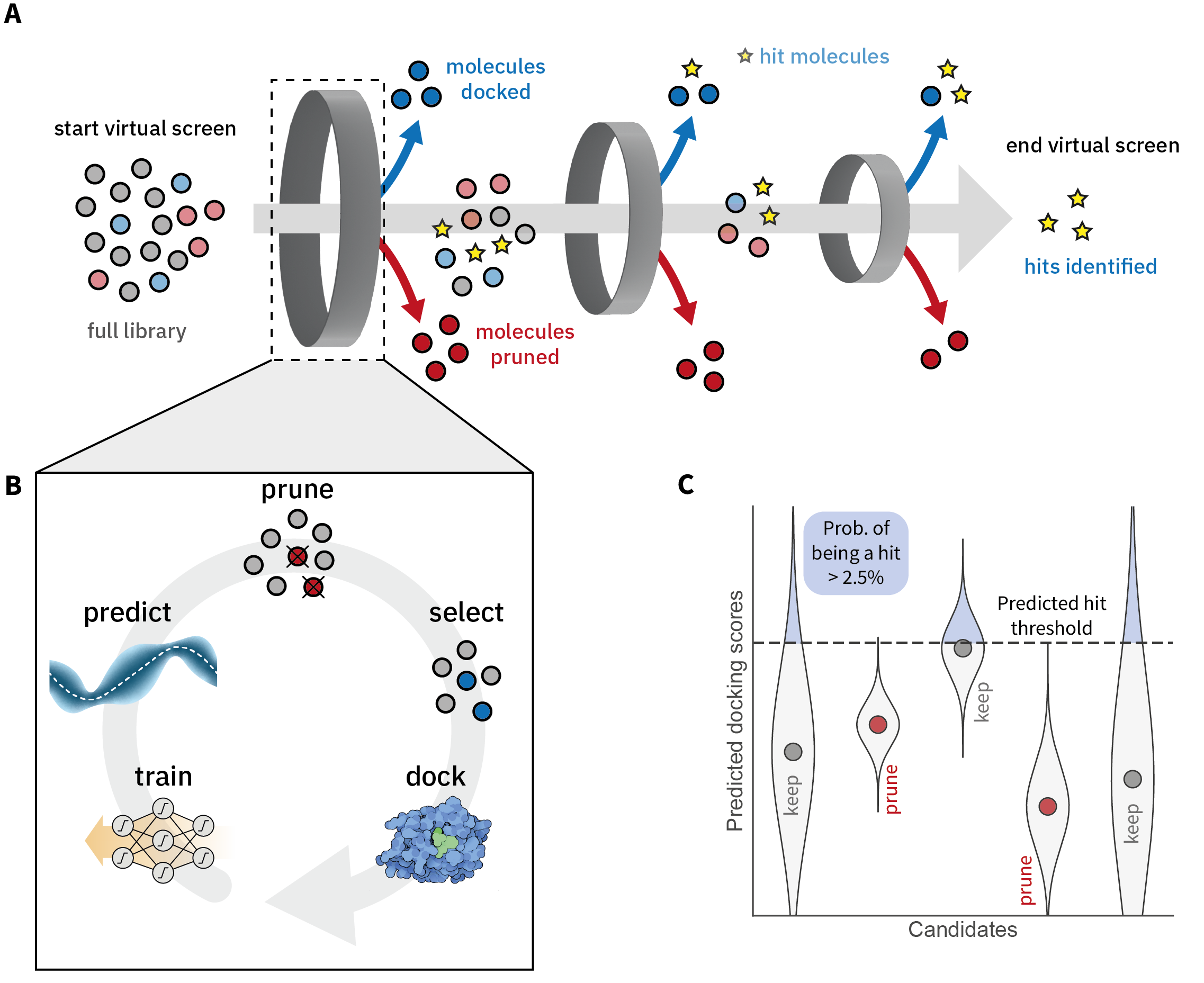}
    \caption{Overview of Bayesian optimization with design space pruning. \textbf{A.} High level view of the virtual screening workflow where compounds are iteratively docked or pruned from the virtual library at each iteration of the optimization loop (grey circle). \textbf{B.} Depiction of a single iteration of the optimization loop. \textbf{C.} Graphical depiction of the pruning step in the optimization loop.}
    \label{fig:intro}
\end{figure}

Though compounds identified through docking may lead to experimentally-validated binders \cite{lyu_ultra-large_2019,gorgulla_open-source_2020}, brute-force screens of these magnitudes require a substantial commitment of computational resources. For example, it took \citeauthor{gorgulla_open-source_2020} 475 CPU-years to screen 1.4B compounds using QVina2\cite{alhossary_fast_2015}, which at the time of publication represented all compounds in the Enamine REAL and Zinc databases. The Enamine REAL database alone now contains more than 21B compounds,\cite{noauthor_real_nodate} so exhaustively screening it would consume over \num{7000} CPU-years worth of resources (assuming the same screening protocol). Moreover, this effort would be limited to targeting a single active site on one receptor, meaning that each new target would require \emph{another} \num{7000} CPU-years worth of resources.

An alternative molecular discovery approach to virtual screening is de novo design.\cite{elton_deep_2019,bilodeau_generative_2022} In this approach, the chemical space is defined implicitly and molecules with desired properties are directly generated rather than identified through a brute-force search.\cite{coley_defining_2021} Despite the promising nature of de novo design, pre-enumerated molecular libraries maintain certain advantages. In particular, synthesizablity remains a challenge in generative models,\cite{gao_synthesizability_2020} and even in the case that proposed molecules are synthesizable, sourcing these molecules is a different challenge. In contrast, searching within a well-defined library, such as catalogs from commercial vendors, reasonably guarantees the availability of a selected compound.

Model-guided optimization strategies enable the efficient search within these predefined libraries. Bayesian optimization (BO) is one such technique that operates by fitting a surrogate machine learning model to observed data and then using this surrogate model to guide the selection of subsequent experiments. Broadly, BO operates via (1) observing some initial data, (2) fitting a surrogate model to these data, (3) using the fitted model to select a new point to observe that maximizes some notion of ``acquisition utility'', (4) observing that point then adding it to the dataset, and (5) repeating steps (2)--(4) for some number of iterations (Algorithm~\ref{algo:dsp}).\cite{frazier_tutorial_2018}
Bayesian optimization has previously been employed in pharmaceutical discovery,\cite{czechtizky_integrated_2013,williams_cheaper_2015,reker_active-learning_2015,pyzer-knapp_bayesian_2018} materials design,\cite{yuan_accelerated_2018,xue_accelerated_2016,balachandran_adaptive_2016} and molecular simulations.\cite{seko_machine_2014,seko_prediction_2015}.
More recent work has also demonstrated its abilities to reduce the costs of identifying the top-performers in HTVS.\cite{gentile_deep_2020,graff_accelerating_2021,yang_efficient_2021,pyzer-knapp_using_2020,ahmed_efficient_2018,svensson_improving_2017,kalliokoski_machine_2021,martin_state_2021,pyzer-knapp_bayesian_2018} 

Yet HTVS remains an underexplored setting in Bayesian optimization (BO) due to both the large and discrete nature of chemical libraries and Gaussian processes (GPs)--the canonical model type used for BO--scaling poorly in both memory and computational time to such large training and inference sets. The use of alternate surrogate models like random forests or neural networks addresses these scalability problems, but nevertheless introduces costs associated with surrogate model training and inference.
At each iteration of the optimization in HTVS, the surrogate model must not only be fit to the acquired data but also predict the performance of \emph{all} remaining candidate molecules to calculate an updated acquisition utility.
When using an oracle function that is relatively inexpensive, such as docking, these costs cannot be neglected. For example, MolPAL \cite{graff_accelerating_2021} spends roughly 9\% and 8\% of its total budget on inference and training costs, respectively, for a hypothetical optimization campaign (Table~\ref{tbl:opt-cost}). This is in contrast to the other applications that use experiments or physics-based simulations as objective functions; the low-data regime of these studies leads to sample count being the only relevant metric in terms of costs.
As vendor catalogs grow ever larger (the latest Enamine REAL contains approximately 21B compounds \cite{noauthor_real_nodate}), inference costs will continue to scale linearly, making even optimization workflows both prohibitively expensive and practically challenging for these virtual libraries; depending on the batch size chosen for iterative optimization, the 9\% inference costs would grow to occupy an even larger fraction of the total cost.

In this work, we propose an extension to the framework of model-guided optimization in discrete design spaces to address these challenges. Our approach, design space pruning (DSP), irreversibly removes subsets of the design space as the optimization proceeds. It incorporates a theoretically justified pruning strategy with a single hyperparameter that controls risk aversion to false negatives (i.e., omission of high-performing compounds). In empirical tests across a wide range of docking tasks, we find that DSP lowers the inference costs of model-guided optimization by around 50\% without significant changes to optimization efficiency or performance. These costs savings become more substantial as the size of virtual libraries used for screening increases. 

\section{Approach}\label{sec:approach}
A central premise of surrogate model-guided optimization is that the iterative acquisition of new information leads to improvements in model performance and guides candidate selection to the best compounds. However, the goals of training a good surrogate and the goals of an optimization are distinct. It is not necessarily true that the data acquired throughout a ``successful'' optimization will lead to a more accurate model. In our experience with docking-based optimization, we observe that while the model becomes progressively better at prioritizing molecules with more desirable docking scores, its accuracy with respect to the overall library changes minimally throughout the optimization (Figures~\ref{fig:hts-sb} and \ref{fig:hts-errors}). That is, molecules that the surrogate model predicts to be poor binders in the first iteration of the optimization generally remain so in subsequent iterations. The computational cost of recalculating acquisition utility is essentially wasted on these unpromising compounds.

To address this issue of wasted inference costs, we introduce a strategy we refer to as design space pruning (DSP). The basic principle of DSP is to irreversibly narrow the design space at each iteration, allowing our model to actively focus its search on progressively more promising regions of space. By shrinking the design space, we reduce the the number of inference calls, and thus the cost, of the acquisition function optimization. The aggressiveness with which compounds predicted to be poor performers are pruned must be a design choice. Unlike other approaches to discrete optimization where we may have strict upper and lower bounds with which to compare two options (e.g., in the branch and bound algorithm),\cite{little_algorithm_1963,land_automatic_1960} here we merely have predicted values and associated uncertainties according to a surrogate model.

In the search for the top-$k$ points in a discrete design space, we define a ``hit'' as a compound with a score at least as good as the $k$\textsuperscript{th} best point, which corresponds to a ``hit threshold'', $y'$. After model training, we predict both the mean, $\hat{\mu}$, and uncertainty, $\hat{\sigma}$, for each point. Analogous to the probability of improvement (PI) acquisition function, we calculate the cumulative distribution function above $y'$ of a normal distribution centered at $\hat{\mu}$ with standard deviation $\hat{\sigma}$ and interpret this value as the probability $p$ that the given point is a hit:
\begin{equation}
    p = \Phi\Big(\frac{\hat{\mu} - y'}{\hat{\sigma}}\Big)
    \label{eqn:pruning}
\end{equation}
Note that this framing presents the optimization objective as a maximization, but we may apply Equation~\ref{eqn:pruning} to a minimization by subtracting the obtained value from 1. We can then make a deterministic decision about whether the given point should be kept or pruned based on the estimated hit probability exceeding some minimum probability threshold, $p^*$. In practice, the true hit threshold used in Equation~\ref{eqn:pruning} is unknown at runtime, but we may approximate it using either the $k$\textsuperscript{th} best observation or the $k$\textsuperscript{th} best prediction $\hat{y}'$.
For all experiments reported in this paper, we set $p^*$ equal to 0.025 and use the $k$\textsuperscript{th} best prediction as our hit threshold. We discuss the implications of using the $k$\textsuperscript{th} best observation in the \ref{sec:variations} section in the supplementary text.
Note that this does not consider the correlation between similar candidates (e.g., similar molecular structures) and treats the predicted scores as independent normal random variables.  When candidates are pruned, they are excluded from the library in subsequent iterations (Figure~\ref{fig:michal}A).

\begin{algorithm}[t!]
    \DontPrintSemicolon
    \KwIn{objective function $f$, acquisition function $\alpha$, surrogate model $\hat{f}$, initial candidate set $\mathcal{X}_0$, hit threshold $k$, minimum hit probability $p^*$}
    Initialize dataset by querying objective function at $n_0$ initial points $D_0 \leftarrow \left\{(x_i,\,f(x_i)) \right\}_{1}^{n_0}$ \;
    \For{$t \leftarrow 1, \ldots, T$}{
        Fit surrogate model $\hat{f}$ using $\mathcal{D}_{t-1}$ \;
        \textcolor{red}{Prune input space $\mathcal{X}_{t}\leftarrow\mathtt{prune}(\mathcal{X}_{t-1},\,\hat{f},\,k,\,p^*)$} \;
        Select $x_t \leftarrow \argmax_{x \in \mathcal{X}_t} \alpha(x; \hat{f},\,\mathcal{D}_{t-1})$ \;
        Update dataset $\mathcal{D}_{t} \leftarrow \mathcal{D}_{t-1} \cup \big\{(x_t,\,f(x_t))\big\}$ \;
    }
    \KwResult{$ {\left\{ x_i \right\}_{i=1}^{k}}^{\mathbf{*}} \coloneqq \argmax_{ \left\{ x_i \right\}_{i=1}^k \subset \mathcal{D}_t } \sum_{i=1}^k f(x_i)$}
    \caption{Bayesian Optimization with Design Space Pruning. Line 4 (red) represents the distinction between BO and BO+DSP.}
    \label{algo:dsp}
\end{algorithm}

This algorithm naturally considers the risk of incorrectly pruning a high-performing candidate by framing the decision in terms of the probability that a molecule's score is better than $\hat y'$. Because the framework of Bayesian optimization already requires the prediction of distributions of  performance (e.g., means and uncertainties) for each candidate in the design space, calculating the CDF requires minimal additional computational effort. Making the pruning decision itself involves setting only one hyperparameter, $p^*$. The full algorithm of BO with DSP may be seen in Algorithm~\ref{algo:dsp}. Therefore, DSP achieves the following desiderata of an ideal model-guided optimization algorithm:
\begin{enumerate}[label=\roman*.,font=\itshape,nosep]
    \item \textit{Control the risk of discarding potentially good molecules.}
    \item \textit{Allow users to tune the level of risk they are willing to incur with minimal hyperparameters.}
    \item \textit{Remain computationally simple without requiring additional predictions beyond Bayesian optimization}
\end{enumerate}

There are some conceptual links between design space pruning and design space \emph{partitioning}, which is a technique that has emerged in high-dimensional BO to decompose the original, high-dimensional optimization into a set of low-dimensional optimizations of lower total complexity than the original problem.\cite{wang_batched_2018,wang_bayesian_2014,wang_learning_2020,eriksson_scalable_2020,groves_efficient_2018} Our approach also shares similar themes to the branch-and-bound algorithm for solving discrete optimization problems.\cite{little_algorithm_1963,land_automatic_1960} At each step of the optimization, we also prune regions of the search space irreversibly; however, unlike branch-and-bound, we don't possess strict upper and lower bounds for each candidate, merely predictions and uncertainties.

The principal motivation of DSP is to lower optimization overhead costs while maintaining overall optimization performance. However, the performance of DSP is bounded by the performance of an optimization that does not prune any inputs. Thus while we may realize benefits in terms of lower overhead costs, the overhead costs saved must be worth the risk of false negatives. Practically, this means that DSP will be useful primarily in settings where overhead costs are non-negligible relative to objective evaluation costs. Examples of this include when (1) the design space is discrete but non-combinatorial, requiring prediction for each individual point at every iteration; (2) the inference costs of the surrogate model are not negligible, whether due to the per-candidate cost or the size of the library; and (3) the objective function $f(\cdot)$ is relatively inexpensive (e.g., can be evaluated in 10's of CPU-seconds, such as computational docking) so that its costs do not overwhelm model costs.
In settings where objective evaluation costs vastly exceed overhead costs, DSP will provide little practical benefit. But when the two costs are comparable, such as BO applied to docking large libraries, DSP can significantly lower the budget expended on non-objective costs as we demonstrate below.

\section{Results and Discussion}

\subsection{Test oracles}
\begin{figure}[t!]
    \centering
    \includegraphics[width=\textwidth]{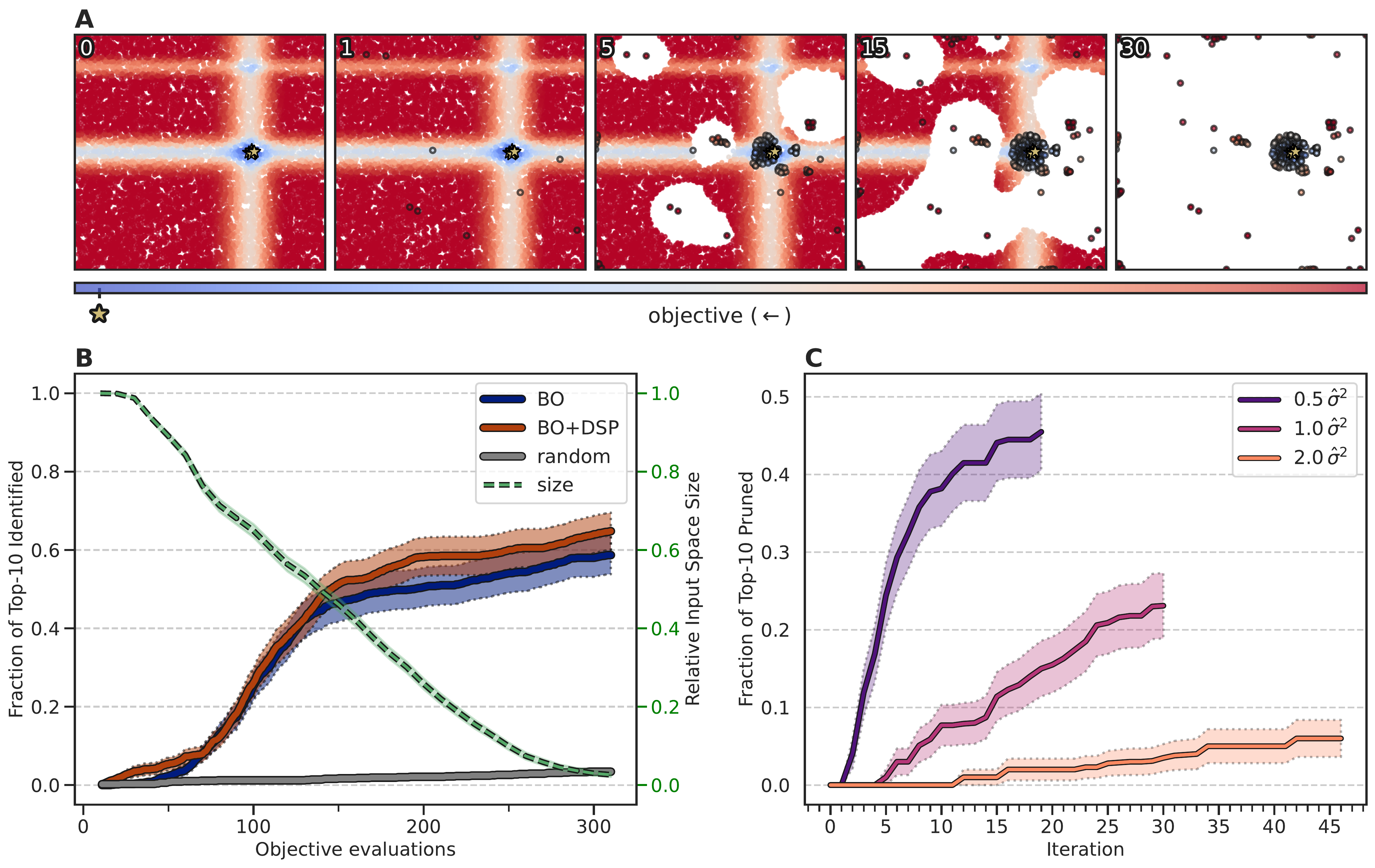}
    \caption{Bayesian optimization with pruning on the discretized Michalewicz function. \textbf{A.} Visualization of design space pruning at the start of the iteration denoted in the upper-left corner of each panel. Acquired points are circled in black. As candidates are pruned, they are removed from the scatterplot. \textbf{B.} Bayesian optimization performance as measured by the fraction of the top-10 points identified (left axis, blue and red traces), and the relative size of the input space when pruning in the optimization (right axis; green, dotted trace). Each trace represents the average $\pm$ standard error of 100 independent runs. \textbf{C.} Fraction of the top-10 points pruned as a function of iteration for when model-predicted variances are scaled up (orange) or down (purple). Each trace represents the average $\pm$ standard error of 100 independent runs.}
    \label{fig:michal}
\end{figure}

We first evaluate design space pruning (DSP) on a suite of optimization benchmark functions that are commonly used in the context of continuous optimization. We adapt them to our setting by discretizing each function into a set of \num{10000} points uniformly sampled from the input domain. Similar to previous work in computational docking optimizations, we frame the task as retrieval of the top-$k$ points from each surface and collect points in batches of size $q$. We construct batches naïvely, selecting the top-$q$ points in the input space ranked by the acquisition function $\alpha$. Additionally, each point may only be queried once during the optimization. We perform each optimization by taking 10 points randomly from the \num{10000} candidate points and then select points in batches of 10 for either 200 iterations or until the design space is ``exhausted'' (i.e., all candidates are either evaluated or pruned). Each optimization was carried out using a Gaussian process (GP) surrogate model with a Matérn-5/2 kernel and an upper confidence bound (UCB) acquisition function.

Applying DSP to the Michalewicz function, we observe that large regions of the design space are pruned by the fifth iteration (Figure~\ref{fig:michal}A). By the 30\textsuperscript{th} iteration of this trial, nearly all of the \num{10000}-point design space has been pruned or acquired, indicating that termination of the optimization is imminent. 
Across the 100 trials, DSP evaluates an average of 325 molecules (minimum of 250 and maximum of 650). In this regime, DSP exhibits nearly identical average performance to the baseline non-pruning approach and identifies slightly more than six of the ten optima on average (Figure~\ref{fig:michal}B). DSP aggressively prunes the design space, with nearly 50\% of the design space eliminated by the 15\textsuperscript{th} acquisition (160 objective evaluations). In turn, this reduces surrogate model inference costs by half, which is further reduced in subsequent iterations. 
Extending our analysis to the maximum sample count, nearly 650 evaluations, DSP finds nearly eight of the ten optima; in contrast, the baseline approach without pruning goes on to find all ten of the optimal points on average (Figure~\ref{fig:all-objs}). 

This performance trend may also be analyzed in terms of the number of top-10 points that were (unfortunately) pruned from the design space (Figure~\ref{fig:michal}C, pink trace). Of the 100 independent runs, pruning generally leads to either the identification of \emph{all} top-10 points in fewer iterations than the baseline or to the identification of \emph{none} of the top-10 because they are pruned prematurely (Figures~\ref{fig:michal-perf-all} and \ref{fig:michal-fpr-all}). A poor random initialization can funnel the initial search towards a local minimum, leading to a model with enough confidence to prune the true optimal region. These results underscore the inherent risks associated with any approach that does not perform an exhaustive search, including the proposed pruning approach despite its strong performance in the majority of trials.

Pruning behavior is also inherently coupled to the predictive uncertainties of our surrogate model. Thus, inflating or deflating these uncertainties should lead to under- or over-pruning, respectively. Multiplying the predicted variance of the GP by 0.5 (deflation) or 2 (inflation) illustrates the expected trends (Figure~\ref{fig:michal}C). Deflated uncertainties lead to more aggressive pruning, earlier termination, and lower average performance, with the reverse being true for inflated uncertainties. High model uncertainty will cause DSP to prune fewer candidates and behave like standard model-guided optimization, whereas overconfident estimates (commonly observed with ensembling) will lead to over-pruning and early termination.

DSP was tested on seven benchmark functions with a variety of landscape characters (Figure~\ref{fig:all-objs}). These results illustrate that DSP produces equal performance to the baseline optimization, showcasing the generality of this approach, and always decreases cost. Functions with clearly defined optimal regions (e.g., Bukin and Six-Hump Camel) lead to more aggressive pruning, while those with relatively flat optimal regions (e.g., Beale, Branin, and Levy) result in more conservative pruning behavior. Exceptions to the general trend of matching performance are the Drop-Wave and Michalewicz functions, where DSP finds an average of 50\% and 80\% of the hits in the design space, respectively, when no limits are placed on the number of oracle calls. We hypothesize that the periodic nature of the Drop-Wave function is ill-suited to the Matérn-5/2 kernel used in these experiments. While DSP exhibits generally strong performance, its performance is doubly sensitive to optimization hyperparameters due to the natural limit pruning imposes on the oracle budget. Though all model-guided optimizations are impacted by the choice of hyperparameters, a poor initialization or surrogate model can lead to premature pruning of the optimal region, permanently bounding the performance of the given trial.

\subsection{Docking with DOCKSTRING}

\begin{figure}[b!]
    \centering
    \includegraphics[width=0.48\textwidth]{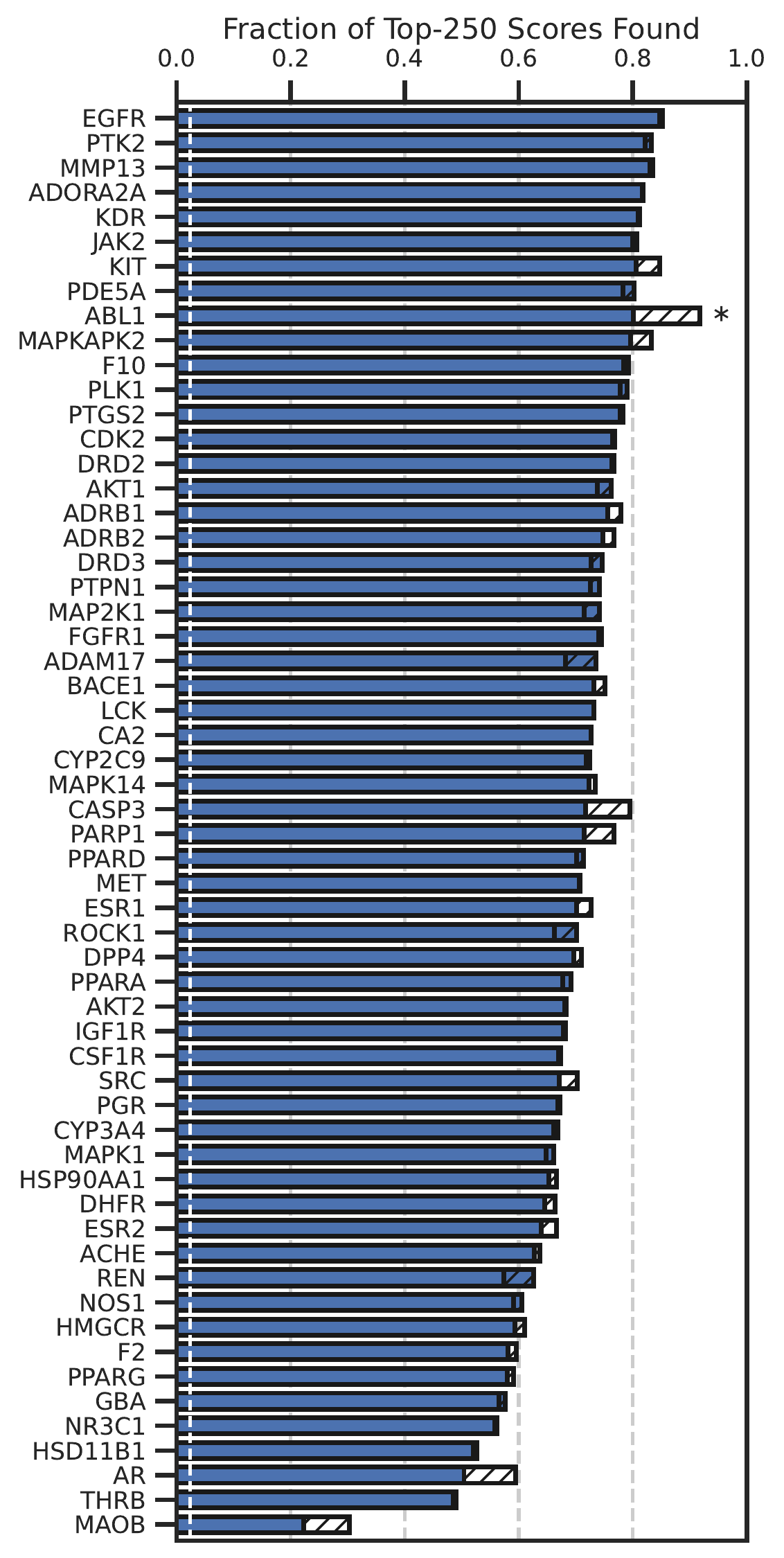}
    \caption{Bayesian optimization performance on DOCKSTRING tasks using DSP, an MPN surrogate model, UCB acquisition function, and 0.4\% batch size. Each bar indicates the average fraction of the top-250 compounds found during five independent runs with docking scores against given target protein as the objective function. Hatched areas indicate the difference in mean performance of the baseline optimization relative to DSP. The white, dotted vertical line represents expected performance of a random search. $\ast$ indicates targets where optimization without pruning outperforms DSP according to a one-sided $t$-test, $p\mathrm{-value}<0.05$ (Bonferroni corrected).}
    \label{fig:dockstring}
    \centering
\end{figure}

As docking is an ideal setting for DSP based on the relative costs of inference and acquisition, we examine the applicability and performance of DSP on 58 docking optimization tasks using the DOCKSTRING benchmark suite,\cite{garcia-ortegon_dockstring_2021} a collection of docking scores of 260k ligands docked against 58 proteins using AutoDock Vina.
We perform retrospective studies on each of the 58 datasets by framing the optimization task as a retrieval of the top-250 (approx. 0.1\%) scores in each dataset. For each optimization, we use a message passing neural network (MPN) surrogate model and select ligands in batches of 0.4\% (approximately \num{1040} ligands per batch) according to their ranking by an upper confidence bound acquisition function ($\beta=2$).\cite{srinivas_information-theoretic_2012}

In 57 of the 58 optimization tasks, there is no statistically significant difference between DSP and the baseline without pruning in terms of the fraction of top-250 compounds they identify after 5 iterations (Figure~\ref{fig:dockstring}). Only for ABL1 do we observe a decrease in performance, with DSP identifying 80\% of the top-250 versus the nearly 92\% found by the baseline. Both DSP and the baseline perform relatively poorly on the MAOB task. The distribution of docking scores for MAOB shows many positive docking scores (approx. \num{4700} of the 260k) and nearly 20 docking scores above $40\ \mathrm{kcal}\cdot\mathrm{mol}^{-1}$ (Figure~\ref{fig:maob-hist}). The presence of these outlier  molecules may lead to a poor surrogate model fit and a worse ability to effectively prioritize molecules relative to the top-250 cutoff. Fortunately, because the surrogate model has high uncertainty for MAOB and other empirically challenging targets (THRB, AR, HDS11B1, and NR3C1), DSP prunes more conservatively than for the ``easier'' targets where model-guided optimization is more successful (Figure~\ref{fig:dockstring-bar+calls-004}).

As intended, DSP significantly reduces overhead optimization costs. Analysis of the cumulative number of inference calls shows that DSP reduces model inference costs by 60-80\% in nearly 90\% of the runs (Figure~\ref{fig:dockstring-calls-004}). We perform similar experiments using both a 0.2\% batch size for nine batches (approx. \num{4700} ligands) and a 0.1\% batch size for 11 batches (approx. \num{2900} ligands). For both of these cases, we observe similar trends as above: DSP massively reduces model inference costs yet achieves a performance that is largely indistinguishable from the no-pruning baseline (Figures~\ref{fig:dockstring-bar-002}-\ref{fig:dockstring-calls-001}).

\subsection{Large-scale docking}
In our previous work applying Bayesian optimization to virtual screening,\cite{graff_accelerating_2021}  we were able to identify roughly 90\% of the top-50k molecules in a 98.2M member library with 40 times fewer simulations than brute force screening, but optimization overhead costs (e.g., model training and inference) required over 15\% of the total workflow budget (Table~\ref{tbl:opt-cost}).

Similar to the DOCKSTRING tasks, we perform a retrospective study on a 98.2M library docked against AmpC $\beta$-lactamase using Glide SP from \citeauthor{yang_efficient_2021},\cite{yang_efficient_2021} framing the task as a retrieval of the top-\num{10000} molecules (comparing scores or SMILES strings, as the two are equivalent in this dataset) according to \citet{yang_efficient_2021}. Molecules are selected in batches of 0.4\% (approx. 400k molecules) using an MPN surrogate model with a UCB acquisition function. The optimization was run for either five batches (excluding the initialization batch) or until the pool of molecules was exhausted.

\begin{figure}[t!]
    \centering
    \includegraphics[width=0.4\textwidth]{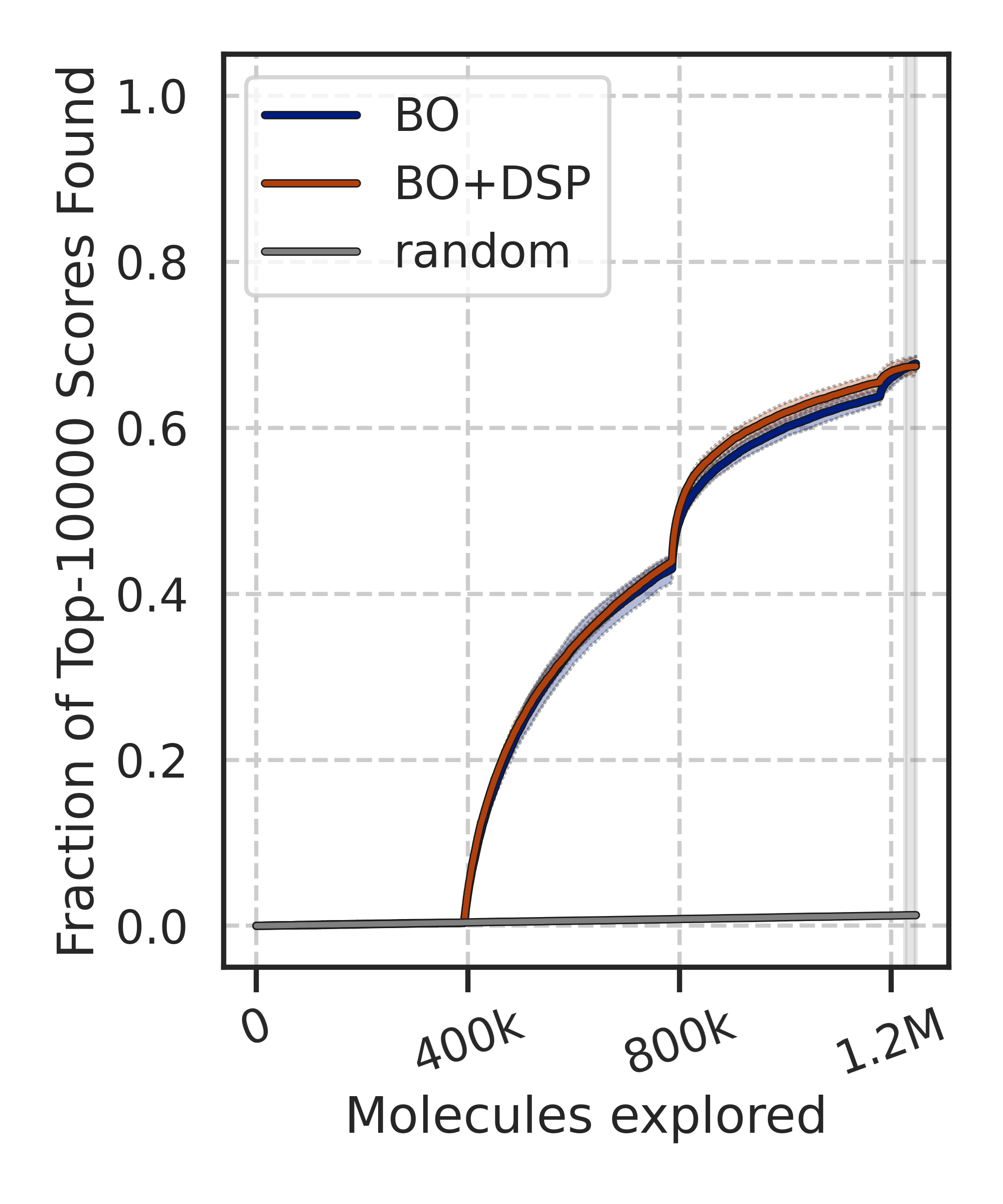}
    \caption{Bayesian Optimization performance on AmpC Glide dataset (98.2M) with an MPN surrogate model, UCB acquisition function, and 0.4\% batch size. The grey shaded area represents the range of possible sample budgets for individual DSP trials. Each trace represents the average $\pm$ standard deviations of three independent runs.} 
    \label{fig:ampc}
\end{figure}

The performance of DSP and the baseline optimization in this ultra large setting are nearly identical (Figure~\ref{fig:ampc}). After approximately 1.2M evaluations, DSP identifies  66.3\%--69.0\% of the top-\num{10000} molecules compared to the 66.0\%--68.8\% that the baseline optimization identifies. Using our estimates of optimization costs, we calculated the total overhead cost (Table~\ref{tbl:opt-cost}) and find that
DSP saves roughly 40\% of overhead costs compared to the baseline approach; in this hypothetical workflow, baseline overheads costs correspond to approximately \$100 of compute on Google Cloud Platform (GCP) whereas DSP costs are only \$60 (Table~\ref{tbl:ampc-cost}). For more details on how these costs were calculated, see the \nameref{sec:opt-costs} section in the supplementary text. This improved efficiency is achieved through DSP's progressive lowering of inference costs at each iteration as more and more candidates are pruned.

We further compare our DSP results on the AmpC Glide task against those of \citeauthor{yang_efficient_2021}, who also investigated the application of active learning to structure-based virtual screening. The authors investigated a variety of batch sizes, ranging from 2\% to 10\% of the total library per batch, and ran each optimization for five iterations.
After two iterations with a 3\% batch size (6.1\% total exploration including a 0.1\% initialization batch, approximately 6M molecules), \citeauthor{yang_efficient_2021} identify roughly 65\% of the top-\num{10000} molecules. In comparison, with the same number of exploration iterations but only 20\% the number of molecules evaluated, we are able to identify the same number of top-\num{10000} molecules while at the same time lowering our inference costs by two thirds (Table~\ref{tbl:ampc-cost}).

\begin{table}[t]
    \small
    \caption{Optimization overhead costs incurred at the given number of molecules sampled for both baseline and DSP approaches. Costs are incurred in steps at the given number of molecules sampled, so any sample count less than \num{392906} would have an overhead cost of \$0}
    \begin{tabular*}{0.55\textwidth}{cl@{\extracolsep{\fill}}l}
        \toprule
        \hfill & \multicolumn{2}{c}{overhead cost / \$} \\ \cmidrule(l){2-3} 
        number of molecules sampled & baseline & DSP \\ \midrule
        \num{392906} & 26.6 & 26.6 \\
        \num{785812} & 59.2 & 39.6 \\
        \num{1178718} & 97.9 & 57.9 \\
        \bottomrule
    \end{tabular*}
    \label{tbl:ampc-cost}
\end{table}

An important element of DSP is its early termination. Model-guided optimization, when run to completion with an infinite budget, would eventually explore the full design space and identify 100\% of the top-\num{10000} molecules. However, with a 0.4\% batch size, DSP ``exhausts'' the design space after exploring only 1.22M--1.24M molecules, i.e., all molecules are either evaluated or pruned at that point. The early termination of DSP, as expected, does lead to false negatives at a rate that is tunable through the hyperparameter $p^*$. Setting this hyperparameter \textit{a priori} is a practical challenge, similar to knowing when to stop a prospective model-guided optimization, . At the time of DSP's natural termination around 1.2M evaluations, DSP identifies more than 67\% of the hits while expending a total compute budget of \$523. Allowing the baseline approach to explore for five full iterations (2.36M evaluations) identifies 82\% of the hits at a total cost of \$1150.

Of further interest is how DSP impacts the search dynamics during the optimization. Because it prunes candidates that the surrogate model believes to be poor-performing molecules, we speculated that certain types of compounds may be more likely to be removed than others, i.e., ``singleton'' compounds that are local optima in the structure-activity landscape. We cluster the top-\num{10000} molecules from the AmpC-Glide dataset using directed sphere exclusion clustering \cite{gobbi_dise_2003,hudson_parameter_1996,butina_unsupervised_1999} and analyze the distribution of cluster sizes in this set (Figure~\ref{fig:ampc-clusters}). Molecules were represented via 2048-bit Morgan fingerprints of radius 2 and clustered using a maximum Tanimoto similarity of 0.35. The distribution of cluster sizes reveals one large cluster with approximately 350 molecules, 292 singleton clusters, and the remaining 9360 molecules residing in clusters of size 2--250 (Figure~\ref{fig:ampc-clusters}).

Using the definitions above, we analyze the overall AmpC performance in terms of how both DSP and the baseline optimization explore molecules belonging to the singleton, mid-sized, or large cluster (Figure~\ref{fig:ampc-004-clusters}). Identifying 100\% of the top-\num{10000} molecules would correspond to identifying 100\% of the singleton, mid, and large cluster molecules. Comparing the rate of recovery for each molecule type by DSP to baseline shows that both approaches explore the space of the top-\num{10000} in a roughly equivalent fashion. This is surprising, as one might expect that DSP would accelerate exploration of molecules in the largest cluster at the expense of singleton molecules. We believe this is due to the pruning function being almost identical in form to the probability of improvement acquisition function. Given a list of candidate molecules ranked by our acquisition function, the optimization will select molecules from the top of this list while the pruning decision is very likely to remove molecules from the bottom of this list. For pruning to materially alter the search dynamics of an optimization, it must prune molecules that would otherwise be selected in subsequent iterations. 

\section{Conclusions}
We propose design space pruning (DSP) as an extension to the framework of surrogate model guided optimization that lowers the additional costs introduced by the surrogate model inference step. Evaluation of DSP on a variety of tasks demonstrates that (1) its ability to identify the best-performing candidates is comparable to that of the baseline optimization and (2) DSP enables significant reductions in overhead costs. This cost consideration is particularly relevant to the task of high-throughput virtual screening using docking, as surrogate model costs are not negligible relative to the docking calculation itself. Applying DSP to an ultra-large docking dataset of 98.2M molecules from \citeauthor{yang_efficient_2021}, we are able to identify 67.7\% of the top-\num{10000} compounds, recapitulating performance of the baseline optimization while reducing overhead costs by 40\%. Our algorithm introduces only one additional hyperparameter that allows a user to balance the risk of pruning true actives with the reduction in inference costs. Additional variants and extensions of this algorithm that could be explored in future work are discussed in the supporting information.

\section{Materials and Methods}

\subsection{Datasets}
Discretized test oracle datasets were generated by drawing \num{10000} points uniformly at random within the bounds of the respective function then evaluating these points with zero noise. DOCKSTRING data was the full benchmark dataset provided in ref. \citenum{garcia-ortegon_dockstring_2021}. AmpC-Glide data was obtained from ref. \citenum{yang_efficient_2021}.

\subsection{Batched Bayesian optimization}
The batched Bayesian optimization procedure was the same for both test oracle and docking datasets. To initialize, $n$ points are selected at random from the discrete design space, evaluated, and added to the dataset. At each iteration, (1) the surrogate model is fit to all available data, holding out 20\% for validation in the case of docking; (2) the trained surrogate model is then used to predict the objective value and uncertainty for each point in the design space; (3) the acquisition utility for each point is calculated; (4) the top-$q$ points ranked by acquisition utility are evaluated with the objective function; (5) the dataset is updated with these new data; and (6) steps (1)--(5) are repeated.

\textbf{Pruning } When DSP is employed, the design space is pruned between steps (2) and (3) from above. Using the predicted objective values and uncertainties along with the estimated hit threshold, the ``hit probability'' of each point is calculated according to Equation~\ref{eqn:pruning}. All points with a calculated value of $p$ less than $p^*$ are then irreversibly pruned from the design space. In all experiments, the value of $p^*$ was set at 0.025.

\subsection{Surrogate models}
\subsubsection{Test oracles}
The optimizations on the test oracles utilized Gaussian process (GP) model with a Matérn-5/2 kernel. GP hyperparameters were optimized over 150 epochs using the Adam optimization algorithm with a learning rate of $1\cdot10^{-3}$ and marginal log likelihood loss function. All code was implemented using the BoTorch package.\cite{balandat_botorch_2020}

\subsubsection{Docking}
The docking-based optimizations (both DOCKSTRING and AmpC-Glide) utilized Chemprop, a message-passing neural network from ref. \citenum{yang_analyzing_2019} and \citenum{hirschfeld_uncertainty_2020}, as the surrogate model. The model was used with the following settings: messages passed on directed bonds, ReLU activation of messages, an encoded dimension of 300, and encoder output fully connected to a feedforward neural network with a single, 300-dimensional hidden layer connected to a 2-dimensional output layer. The model was trained over 50 epochs with early stopping using Adam optimization, a Noam learning rate scheduler (initial, maximum, and final learning rates of $1\cdot10^{-4}$, $1\cdot10^{-3}$, and $1\cdot10^{-4}$, respectively, warming up over two epochs), and MVE loss from ref. \citenum{nix_estimating_1994}. Early stopping tracked the validation loss and had a patience of 10 epochs. The surrogate model was trained fully from scratch at each iteration of the optimization with no hyperparameter tuning.

\subsection{Evaluation}
Optimization performance was evaluated as the retrieval of the top-$k$ scores in the design space. This is similar, in principle, to the top-$k$ molecules but it accounts for the fact that multiple molecules may have equal scores. This metric is calculated by taking the intersection of the list of true top-$k$ scores and observed top-$k$ scores divided by $k$.

\section*{Author Contributions}
D.E.G. authored the software and performed the experiments. D.E.G. and M.A. designed the figures. D.E.G., M.A., and C.W.C. conceptualized the idea. D.E.G., M.A., and C.W.C. wrote the original draft. E.P.K, K.E.J, and C.W.C. supervised the work. All authors contributed to the analysis of results and the final version of the manuscript.

\section{Data and Software Availability}
All code needed to reproduce the data in this study and the resulting analysis can be found at \url{https://github.com/coleygroup/molpal} (docking-based oracle functions) and \url{https://github.com/coleygroup/dsp} (toy oracle functions).

\FloatBarrier
\begin{acknowledgement}
The authors thank Wendy Cornell for useful discussions to shape the direction of this work and for commenting on the manuscript. This work was funded by the MIT-IBM Watson AI Lab. The authors acknowledge the MIT SuperCloud and Lincoln Laboratory Supercomputing Center for providing HPC resources that have contributed to the research results reported within this paper.
\end{acknowledgement}

\begin{suppinfo}
Additional methods and results can be found in the supporting information.
\end{suppinfo}

\bibliography{refs}

\clearpage
\input{supp}
\clearpage

\end{document}

%% file: supp.tex
\renewcommand{\thefigure}{S\arabic{figure}}
\renewcommand{\thetable}{S\arabic{table}}
\setcounter{figure}{0} 
\setcounter{table}{0} 
\setcounter{page}{1}

\begin{center}
    \sffamily\bfseries
    \huge{Supporting Information} \\
    \LARGE{Self-focusing virtual screening with active design space pruning} \\
    \vspace{0.75em}
    \mdseries
    \large{
        David E. Graff,$^{\dagger,\ddagger}$
        Matteo Aldeghi,$^{\ddagger}$
        Joseph A. Morrone,$^{\P}$
        Kirk E. Jordan,$^{\S}$
        Edward O. Pyzer-Knapp,$^{\parallel}$
        and Connor W. Coley$^{\ast,\ddagger,\perp}$
    } \\
    \vspace{0.75em}
    \normalsize\rmfamily\itshape
    $\dagger$~Department of Chemistry and Chemical Biology, Harvard University, Cambridge, MA 02138 \\
    \vspace{0.1em}
    $\ddagger$~Department of Chemical Engineering, MIT, Cambridge, MA 02142 \\
    \vspace{0.1em}
    \textup{\P}~Computational Biology Center, IBM Thomas J. Watson Research Center, Yorktown Heights, NY 10594 \\
    \vspace{0.1em}
    \textup{\S}~IBM Thomas J. Watson Research Center, Cambridge, Massachusetts 02142 \\
    \vspace{0.1em}
    $\parallel$~IBM Research Europe, Hartree Centre, Daresbury WA4, U.K. \\
    \vspace{0.1em}
    $\perp$~Department of Electrical Engineering and Computer Science, MIT, Cambridge, MA 02142 \\
    \vspace{0.75em}
    \normalfont
    \textsf{E-mail: ccoley@mit.edu}
\end{center}

\section{Additional Methods}
\subsection*{Variants of the proposed algorithm}\label{sec:variations}
There are several variations of the DSP algorithm proposed in the main text, and we discuss a few of them below.

\begin{figure}
    \centering
    \includegraphics[width=0.4\textwidth]{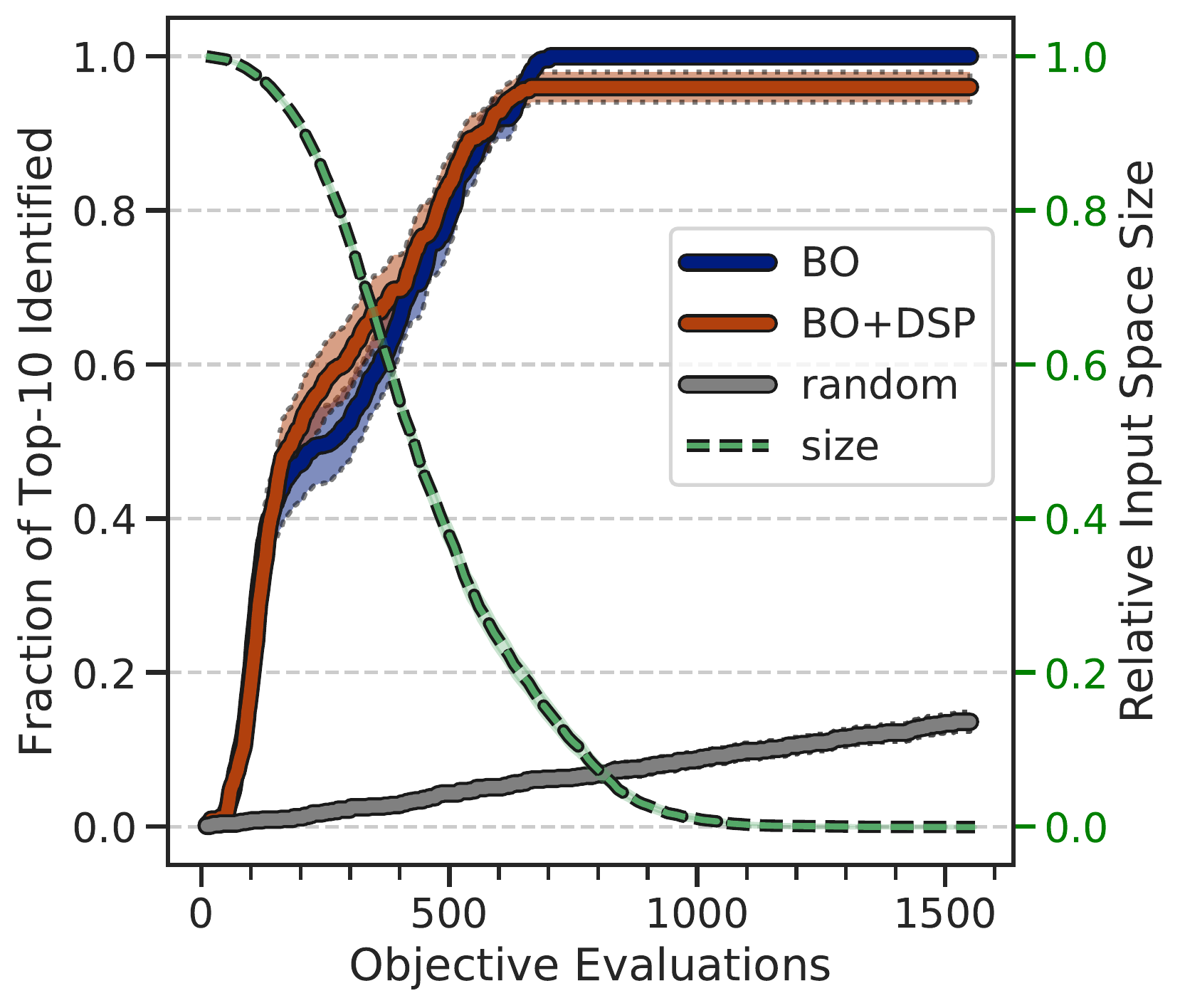}
    \caption{Fraction of the top-10 points identified on the Michalewicz function and relative input space remaining versus objective evaluations when using the \emph{observed} hit threshold for the pruning decision. Each trace represents the average $\pm$ standard error of 100 independent runs.}
    \label{fig:michal-obs}
\end{figure}

\paragraph{Observed threshold} As mentioned in the \nameref{sec:approach} section of the main text, the true of value of the hit threshold $y^*$ is unknown at runtime. All experiments in the main text used the $k$\textsuperscript{th} best prediction, $\hat{y}^*$, to approximate this value, but it is also possible to use the $k$\textsuperscript{th} best observation. In practice, this value will generally be lower than $\hat{y}^*$, leading to larger calculated hit probabilities across the design space (Equation~\ref{eqn:pruning}). Using the observed threshold will thus result in more conservative pruning and a lower risk of removing true hits from the design space over the course of the optimization. We tested this approach on the Michalewicz function and found that it indeed led to more conservative pruning compared to using $\hat{y}^*$ and performance more comparable to the baseline optimization in the limit of larger sample counts (Figure~\ref{fig:michal-obs}).

\paragraph{Pruning once} An alternative to pruning in every iteration is to only prune in the first iteration of the optimization. This is a more conservative approach than the standard, iterative pruning and will understandably lead to lower likelihood of performance loss at the cost of increased inference calls. Depending on the optimization task, this approach can be almost indistinguishable from the baseline approach. For example, the test oracle tasks exhibit minimal pruning in the first iteration. In contrast, the docking tasks typically pruned about half of the design space after the first iteration.

\paragraph{$p^*$ scheduling} It is also possible to schedule the value of $p^*$ used in the pruning decision to more conservatively prune in the beginning when the model has less data and more aggressively as the optimization proceeds. As there a number of ways in which to schedule this value, we leave the investigation of this idea to future work.

\begin{figure}
    \centering
    \includegraphics[width=0.4\textwidth]{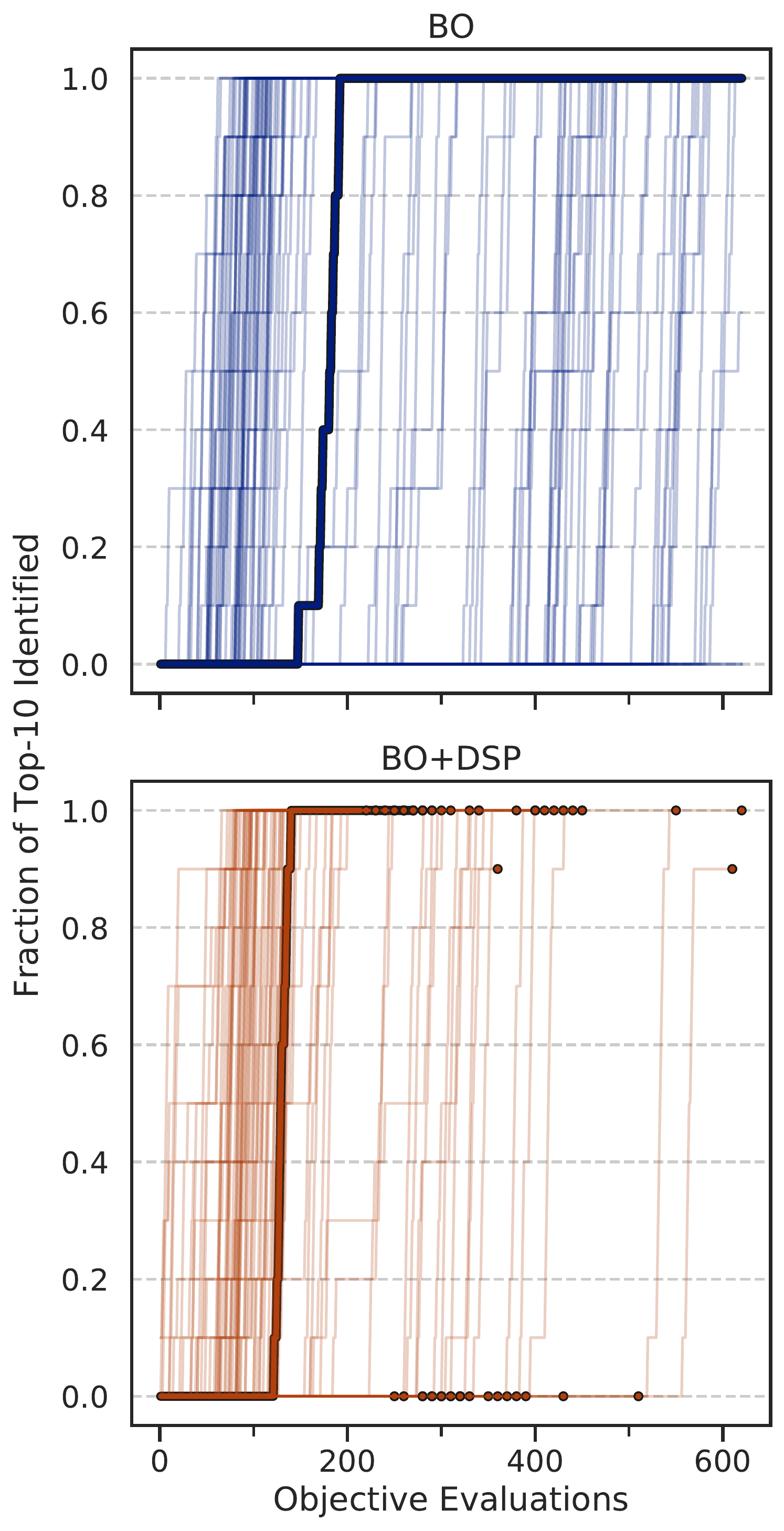}
    \caption{Fraction of the top-10 points identified on the Michalewicz function versus objective evaluations for each trial of both baseline and DSP optimization. The median trial is outlined. A circle signifies the point at which a DSP trial terminates.}
    \label{fig:michal-perf-all}
\end{figure}

\begin{figure}
    \centering
    \includegraphics[width=0.4\textwidth]{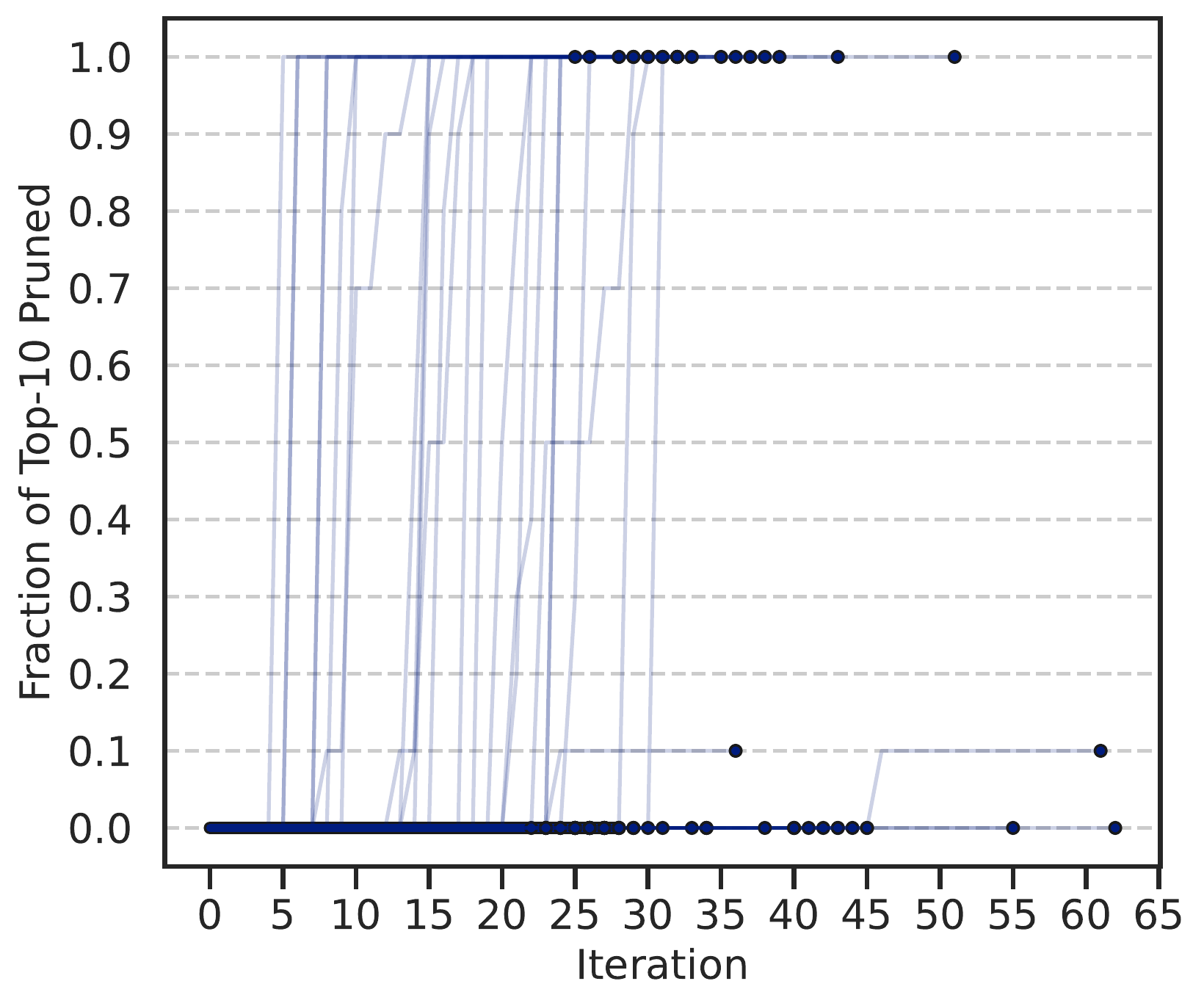}
    \caption{Fraction of the top-10 points falsely pruned on the Michalewicz function versus iteration for each trial. The median trial is outlined. A circle signifies the point at which a trial terminates.}
    \label{fig:michal-fpr-all}
\end{figure}

\paragraph{Initialization} Optimization performance is dependent on initialization conditions and DSP is doubly sensitive. As exemplified in Figures~\ref{fig:michal-perf-all} and \ref{fig:michal-fpr-all}, a poor initialization can funnel the search towards local optima and eventually prune the true optima. Careful initialization strategies can help mitigate this possibility, but investigation of the many possibilities was beyond the scope of this study.

\subsection*{Optimization costs}\label{sec:opt-costs}
We calculated the per-step costs of our optimization workflow on the AmpC-Glide dataset using our observed wall times and Google Cloud Platform (GCP) resource costs (Tables~\ref{tbl:gcp-cost} and \ref{tbl:opt-cost}). We make the following assumptions when calculating GCP costs: we are optimizing within the same virtual library (98.2M molecules), a 0.4\% batch size (approx. 400k molecules per batch), no early stopping during training, our objective function is a Glide docking simulation (30 s/molecule according to ref. \citenum{yang_efficient_2021}), the costs of both the acquisition function calculation and selection steps are negligible, and that wall-times observed on our machines will be the same as those on GCP. Both the docking and inference costs are constant at each iteration, but surrogate model training costs scale linearly with with the size of the dataset, approximately \$6.20 for every 400k molecules in the training set.

\begin{table}
    \centering
    \begin{threeparttable}
    \begin{tabular}{@{}lr@{}}
    \toprule
    resource & cost \\ \midrule
    1 $\mathrm{CPU}\cdot\mathrm{h}$ & \$0.048 \\
    1 $\mathrm{GPU}\cdot\mathrm{h}$\tnote{a} & \$2.48 \\ \bottomrule
    \end{tabular}
    \begin{tablenotes}
    \item[a] Nvidia V100
    \end{tablenotes}
    \end{threeparttable}
    \caption{resource costs on GCP N1 standard machines (accessed 08/2021)}
    \label{tbl:gcp-cost}
\end{table}
\begin{table}
    \centering
    \begin{threeparttable}
    \begin{tabular}{@{}lrrr@{}}
    \toprule
    step & CPU wall time & GPU wall time & GCP N1 cost \\ \midrule
    training\tnote{a} & $7.5 \cdot 10^4$ $\mathrm{CPU}\cdot\mathrm{s}$ & $7.5 \cdot 10^3$ $\mathrm{GPU}\cdot\mathrm{s}$ & \$6.2 \\
    inference\tnote{b} & $2.5 \cdot 10^5$ $\mathrm{CPU}\cdot\mathrm{s}$ & $2.5 \cdot 10^4$ $\mathrm{GPU}\cdot\mathrm{s}$ & \$20.5 \\
    docking\tnote{c} & $1.2 \cdot 10^7$ $\mathrm{CPU}\cdot\mathrm{s}$ & \textit{n/a} & \$158.3\\ \bottomrule
    \end{tabular}
    \begin{tablenotes}
    \item[a] 50 epochs of MPN training with 400k molecules
    \item[b] MPN inference over 98.2M molecules
    \item[c] docking 400k molecules
    \end{tablenotes}
    \end{threeparttable}
    \caption{Observed wall times and calculated costs on GCP N1 virtual machines for the specified optimization steps. Wall times were for MPN training and inference were measured using 10 Intel 6248 CPUs and 1 Nvidia V100 GPU. The wall time for computational docking varies greatly with simulation parameters, thus the values provided are only estimates.}
    \label{tbl:opt-cost}
\end{table}

\FloatBarrier
\section{Additional Results}

\begin{figure}
    \centering
    \includegraphics[width=0.5\textwidth]{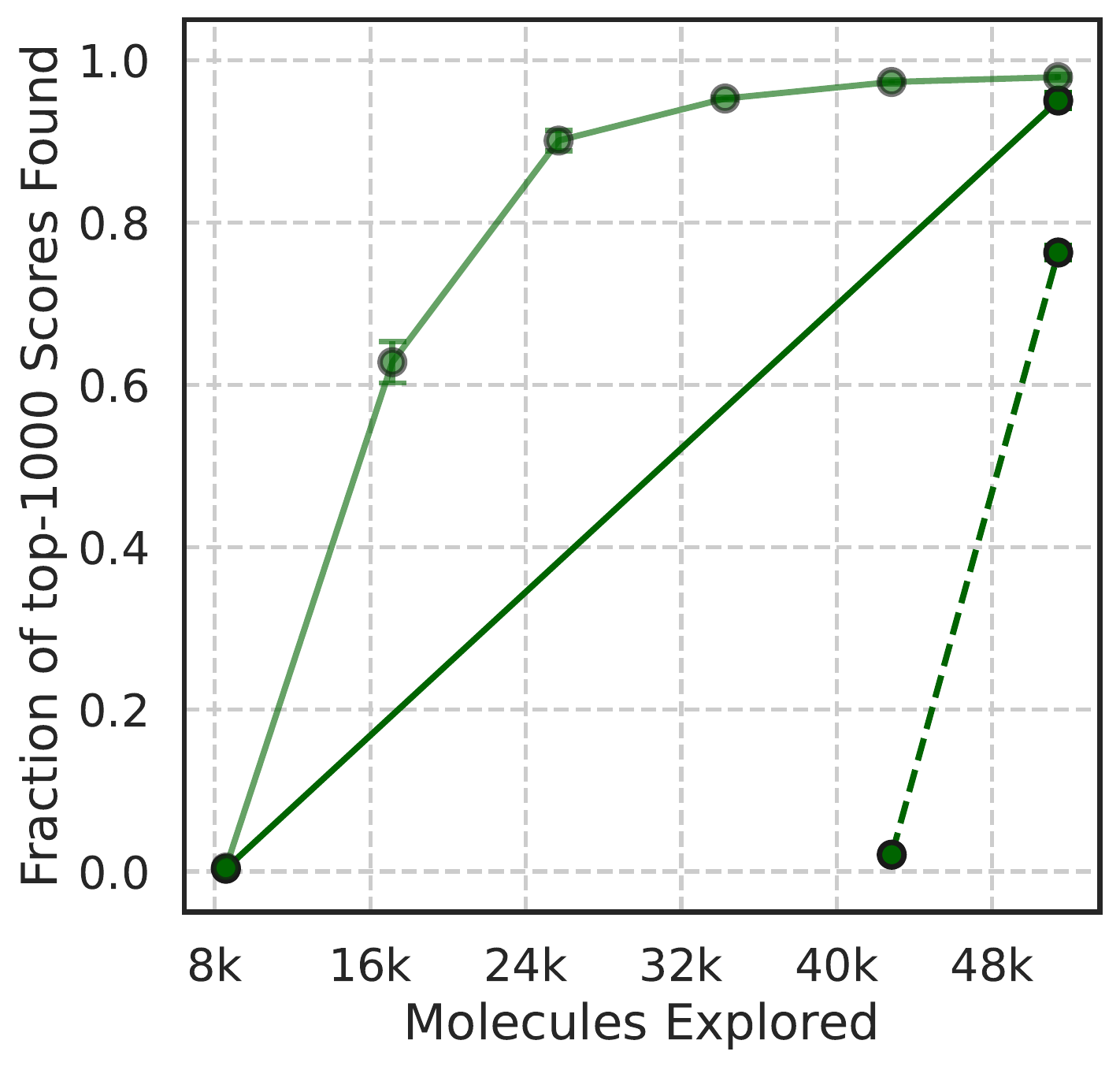}
    \caption{Bayesian optimization performance on the Enamine HTS dataset using an MPN surrogate model, greedy acquisition, and 0.4\% initialization batch. Faded trace: active learning with a 0.4\% batch size. Solid trace: single-batch acquisition with a 2\% acquisition batch. Dotted trace: single-batch acquisition with a 2\% initialization batch and 0.4\% acquisition batch. Figure reproduced from data reported in ref. \citenum{graff_accelerating_2021}.}
    \label{fig:hts-sb}
\end{figure}

\begin{figure}
    \centering
    \includegraphics[width=0.9\textwidth]{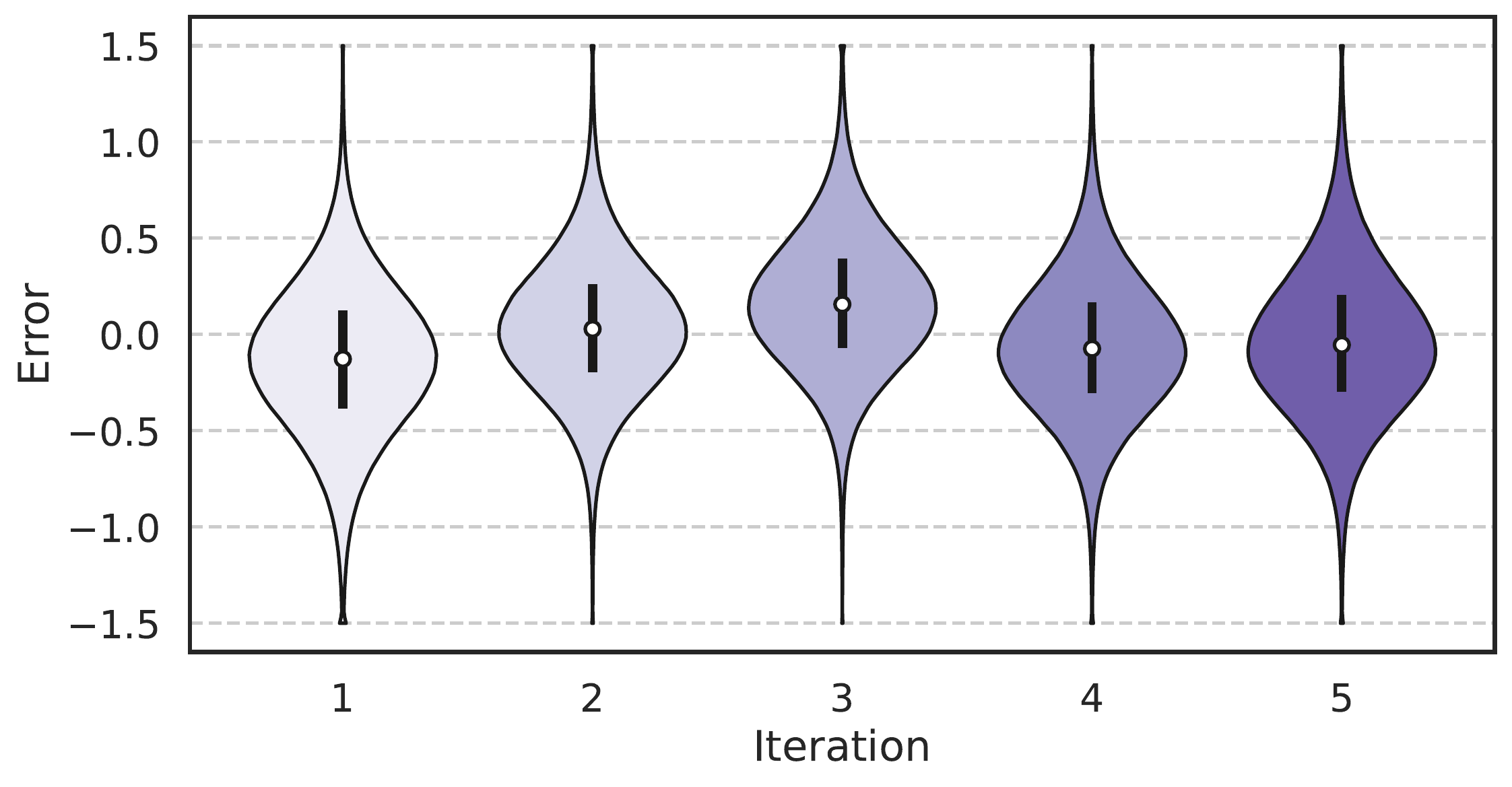}
    \caption{Predictive error of the surrogate model on the full Enamine HTS dataset at the start of the given iteration for active learning using an MPN surrogate model, greedy acquisition, and a 0.4\% batch size. Errors are clipped to the range [-1.5, 1.5] for ease of visualization. White dots represent the median error and black bars represent the Q1--Q3 range. Figure reproduced from data reported in ref. \citenum{graff_accelerating_2021}.}
    \label{fig:hts-errors}
\end{figure}

\begin{sidewaysfigure}
    \centering
    \includegraphics[width=\textheight]{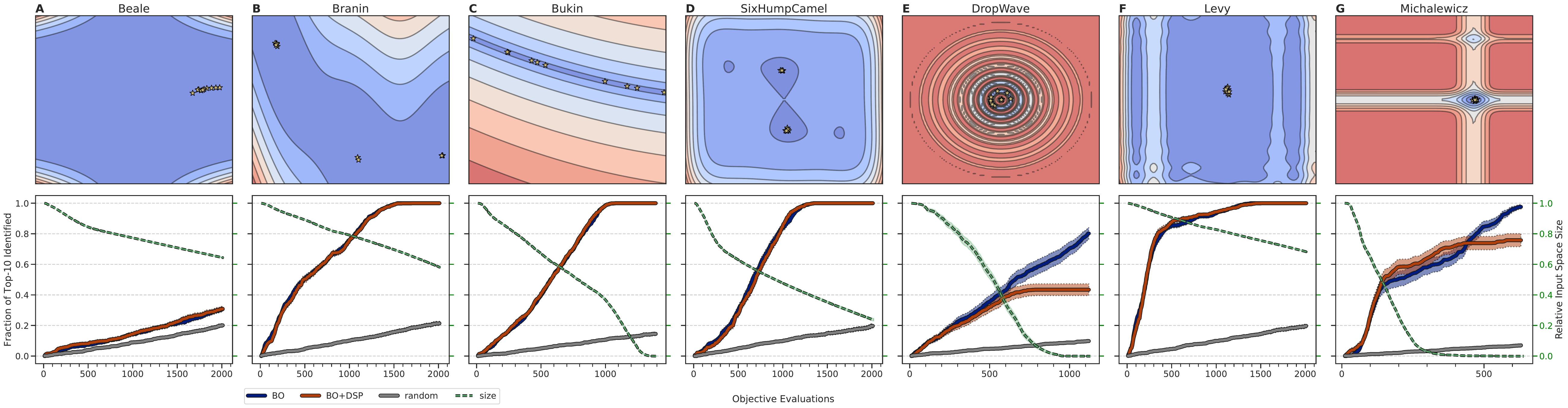}
    \caption{\textbf{Top.} Contour plots for the (continuous) specified function. \textbf{Bottom.} Fraction of the top-10 points identified on the given function and relative input space remaining versus objective evaluations. Each trace represents the average $\pm$ standard error of 100 independent runs.}
    \label{fig:all-objs}
\end{sidewaysfigure}

\begin{figure}
    \centering
    \includegraphics[width=\textwidth]{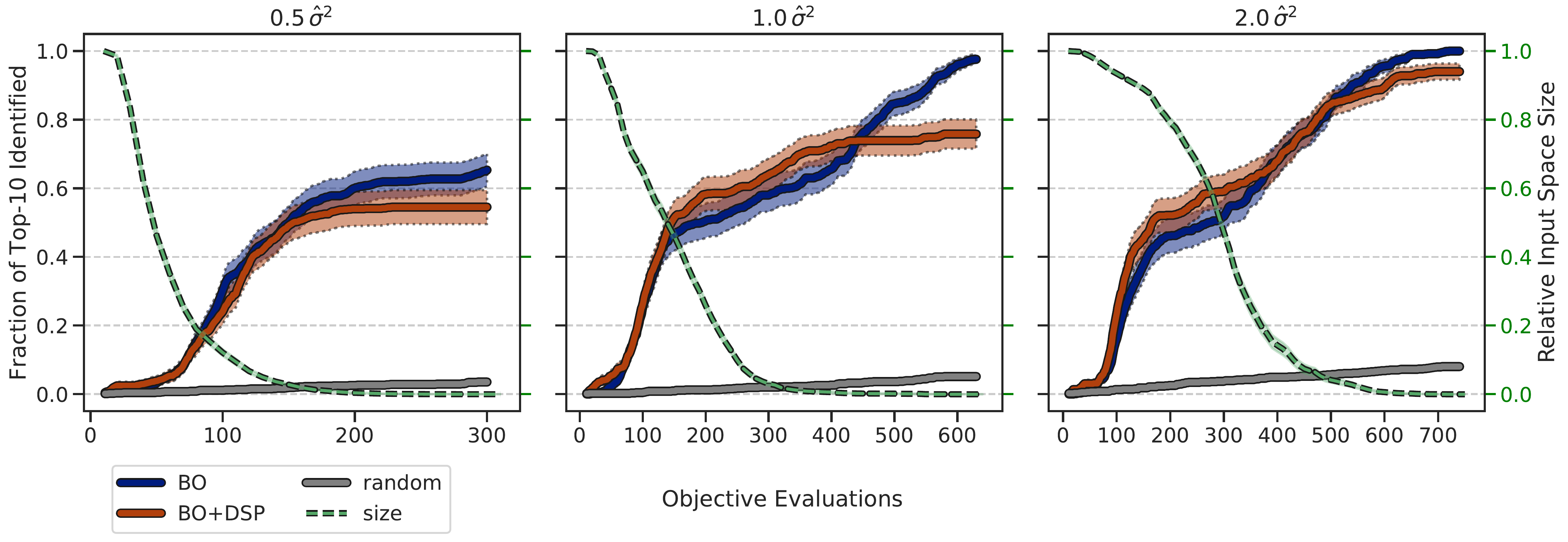}
    \caption{Fraction of the top-10 points identified on the Michalewicz function and relative input space remaining versus objective evaluations when uncertainties provided to the pruning function are multiplied by the given values. Each trace represents the average $\pm$ standard error of 100 independent runs.}
    \label{fig:michal-gamma}
\end{figure}

\FloatBarrier

\begin{figure}
    \centering
    \includegraphics[height=0.8\textheight]{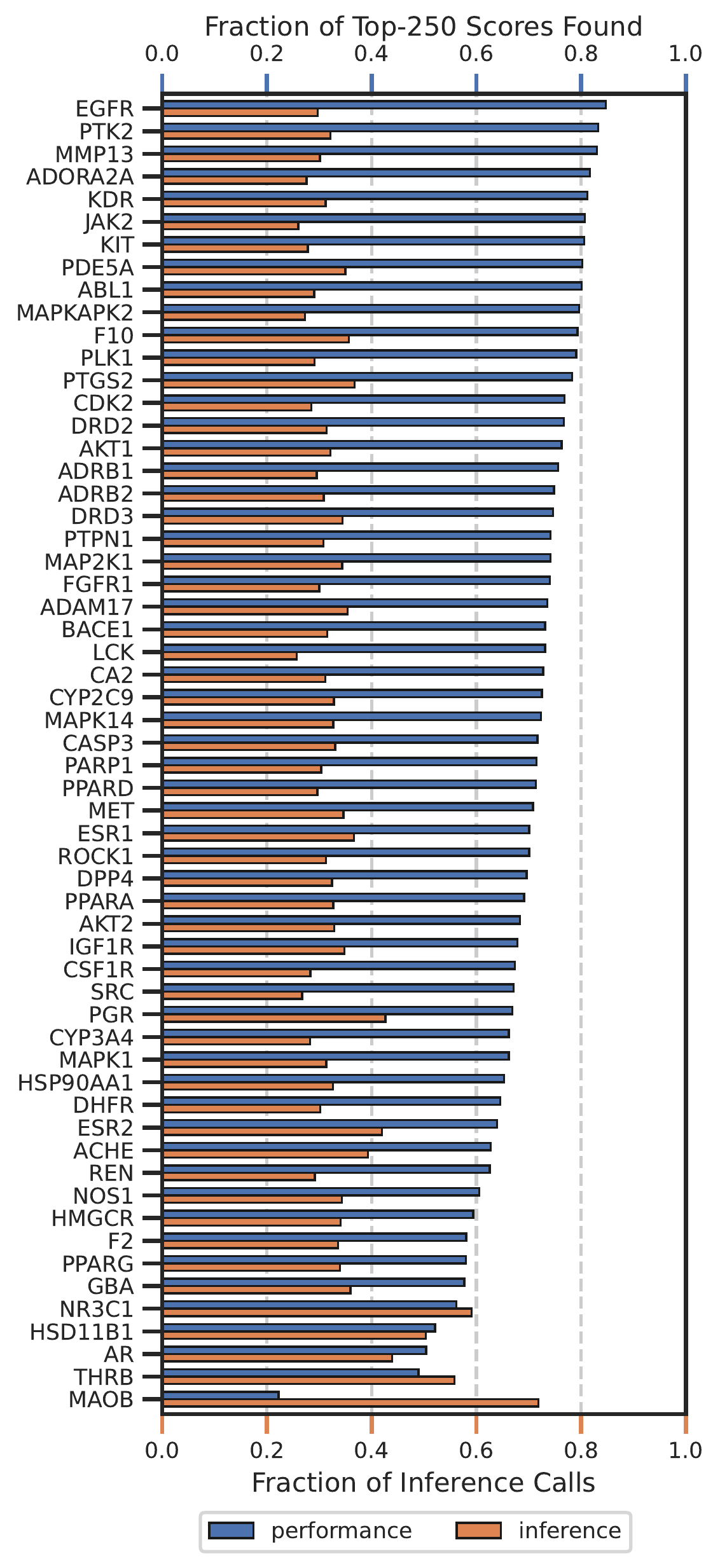}
    \caption{Bayesian optimization performance using DSP (top axis) and total number of inference calls using DSP relative to baseline optimization on DOCKSTRING tasks using a 0.4\% batch size over five iterations. Each bar represents the mean of five independent runs.}
    \label{fig:dockstring-bar+calls-004}
\end{figure}

\begin{figure}
    \centering
    \includegraphics[width=0.6\textwidth]{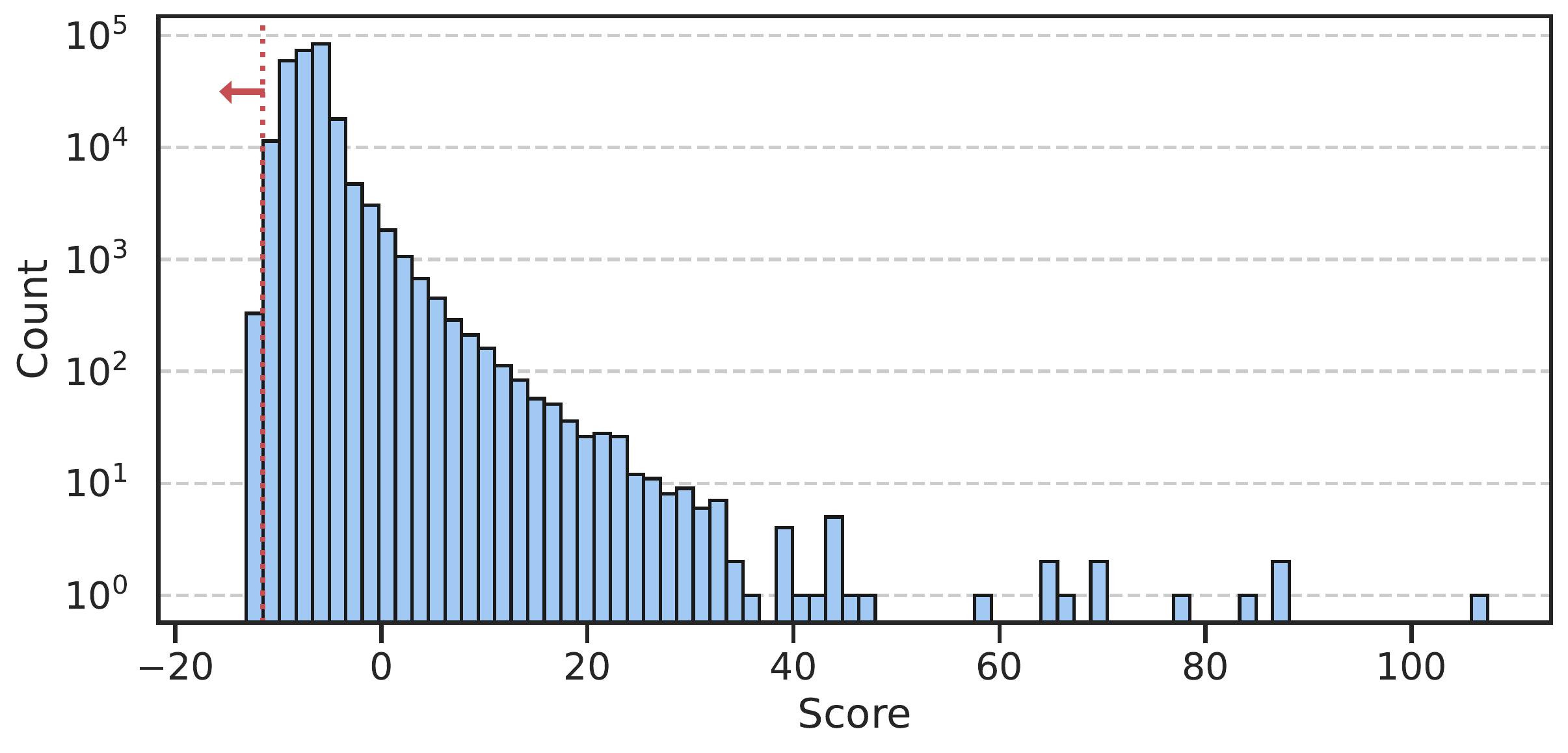}
    \caption{Histogram of scores in the MAOB dataset from DOCKSTRING. Red, dotted line shows the top-250 threshold and red arrow indicates direction of better scores}
    \label{fig:maob-hist}
\end{figure}

\begin{figure}
    \centering
    \includegraphics[width=0.6\textwidth]{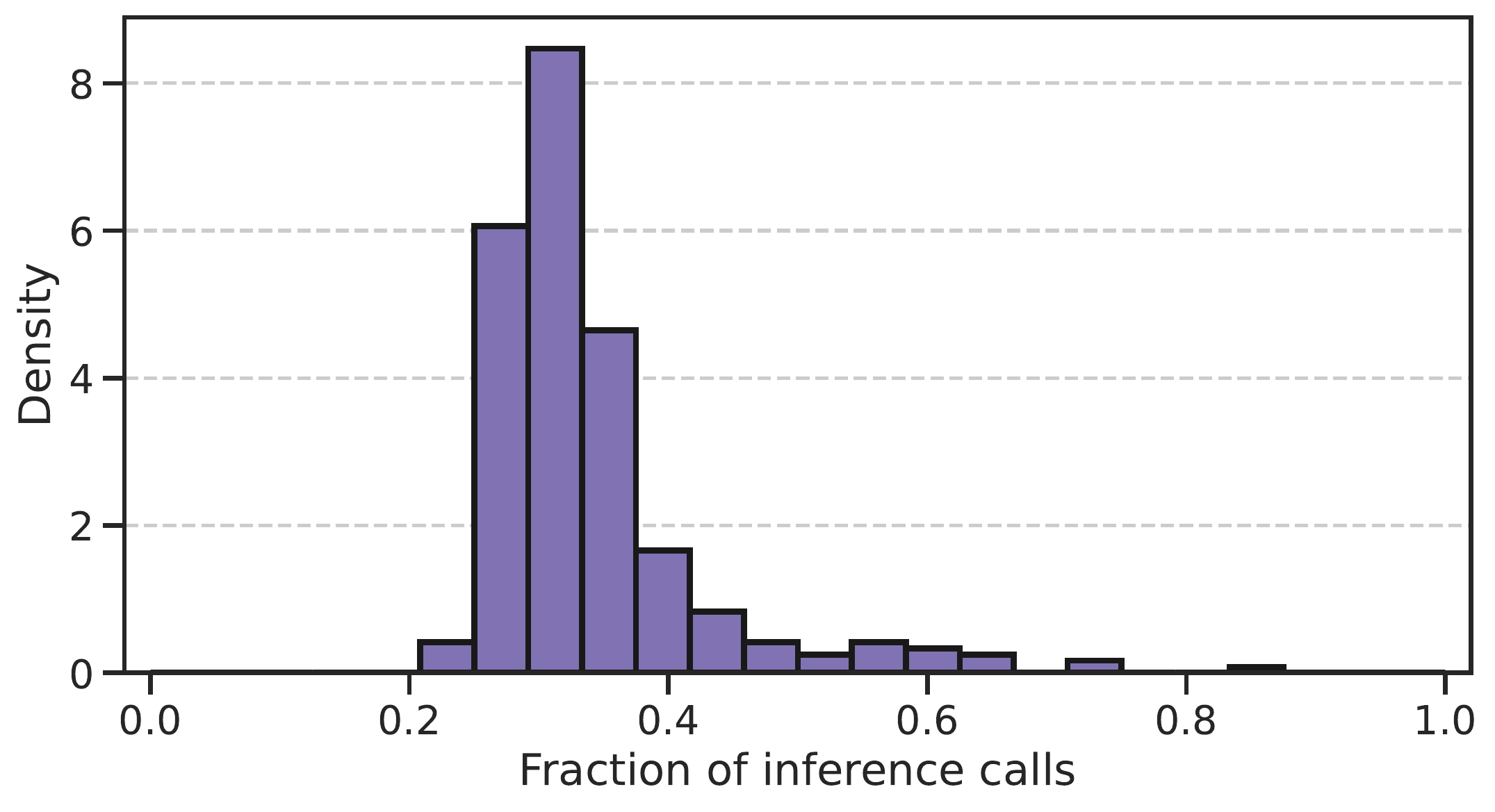}
    \caption{Histogram of the total number of inference calls made using DSP relative to the baseline optimization for a 0.4\% batch size over five iterations.}
    \label{fig:dockstring-calls-004}
\end{figure}

\begin{figure}
    \centering
    \includegraphics[height=0.8\textheight]{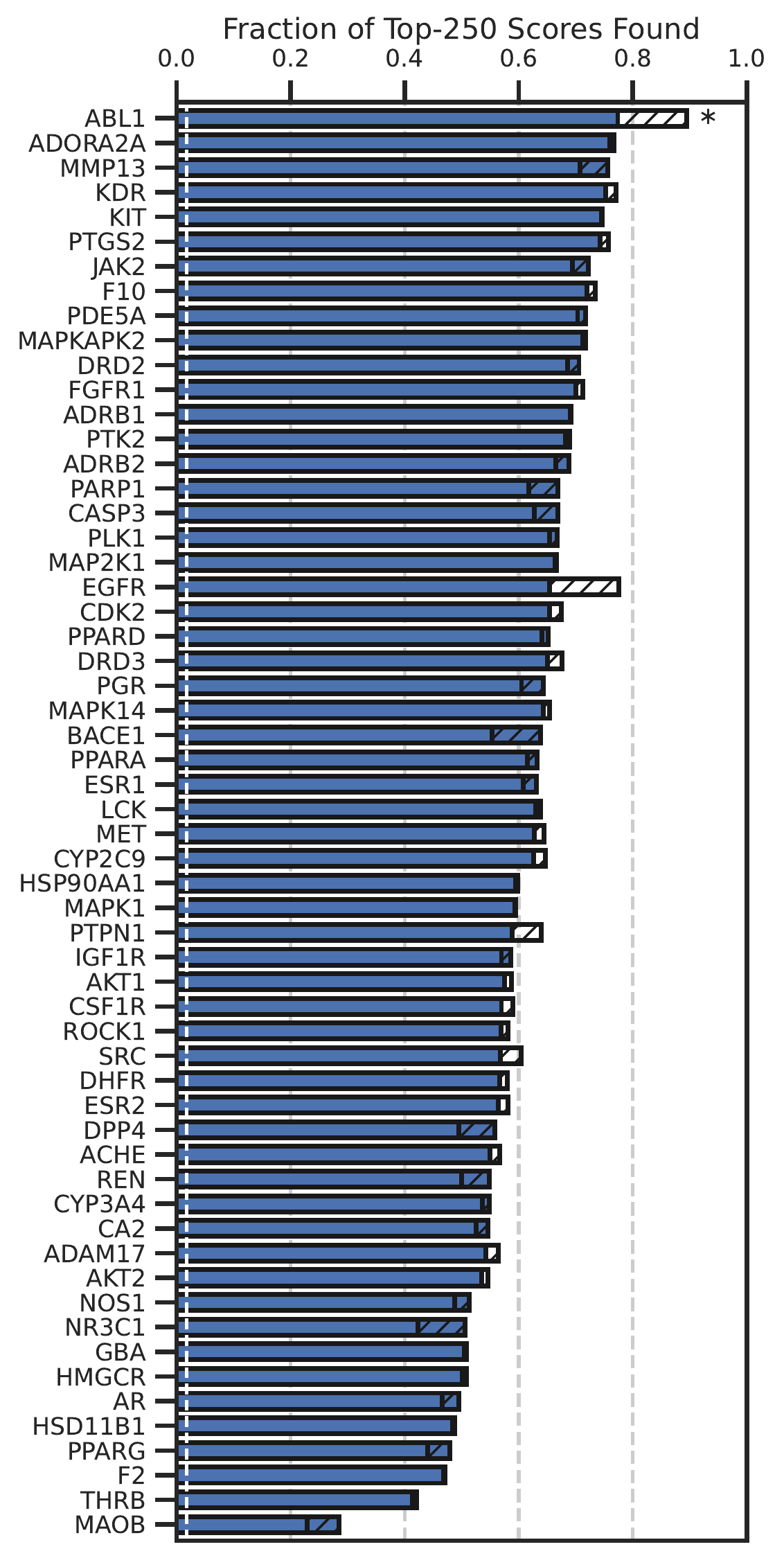}
    \caption{Bayesian optimization performance on DOCKSTRING tasks using DSP, an MPN surrogate model, UCB acquisition function, and 0.2\% batch size. Each bar indicates the average fraction of the top-250 compounds found during five independent runs with docking scores against given target protein as the objective function. Hatched areas indicate the difference in mean performance of the baseline optimization relative to DSP. The white, dotted vertical line represents expected performance of a random search. $\ast$ indicates targets where optimization without pruning outperforms DSP according to a one-sided $t$-test, $p\mathrm{-value}<0.05$ (Bonferroni corrected).}
    \label{fig:dockstring-bar-002}
\end{figure}

\begin{figure}
    \centering
    \includegraphics[height=0.8\textheight]{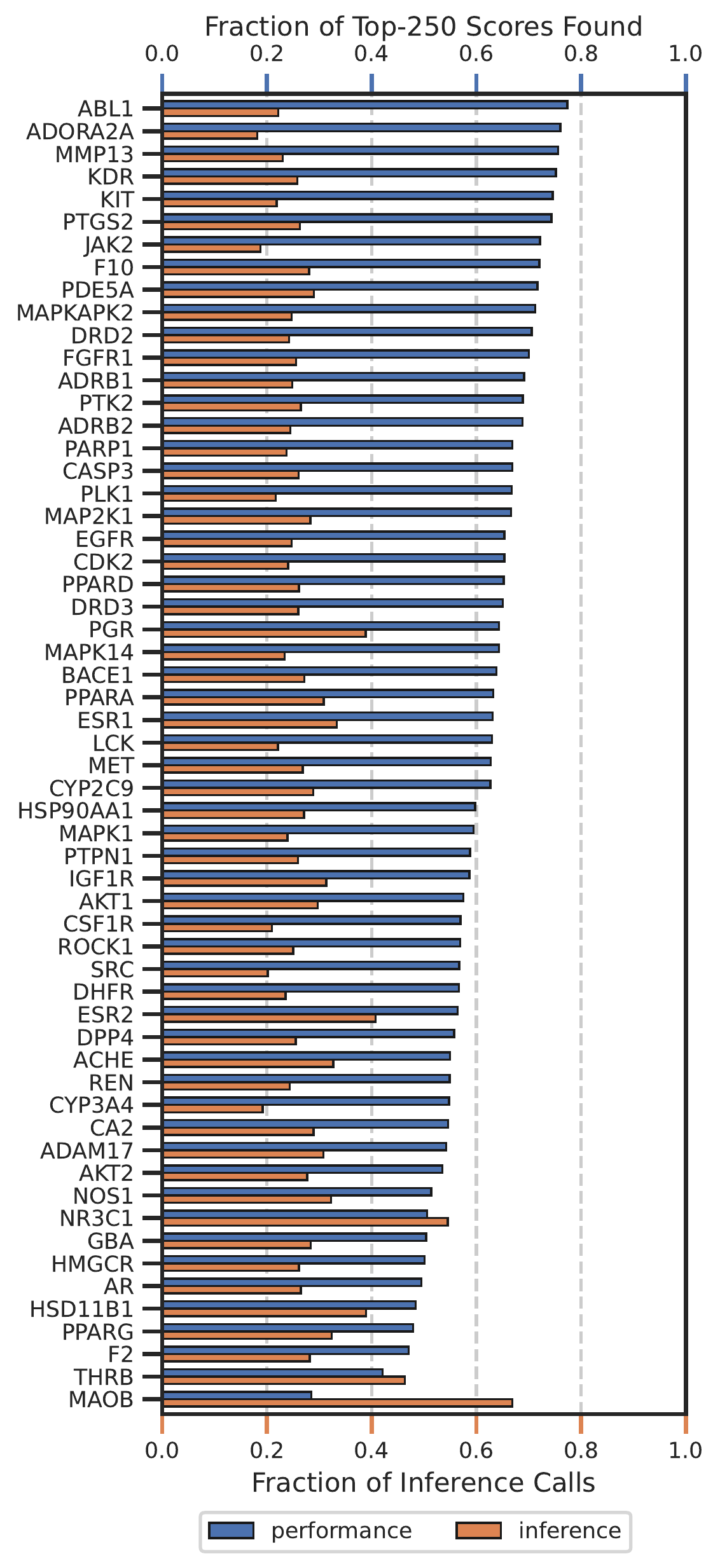}
    \caption{Bayesian optimization performance using DSP (top axis) and total number of inference calls using DSP relative to baseline optimization on DOCKSTRING tasks using a 0.2\% batch size over eight iterations. Each bar represents the mean of five independent runs.}
    \label{fig:dockstring-bar+calls-002}
\end{figure}

\begin{figure}
    \centering
    \includegraphics[width=0.6\textwidth]{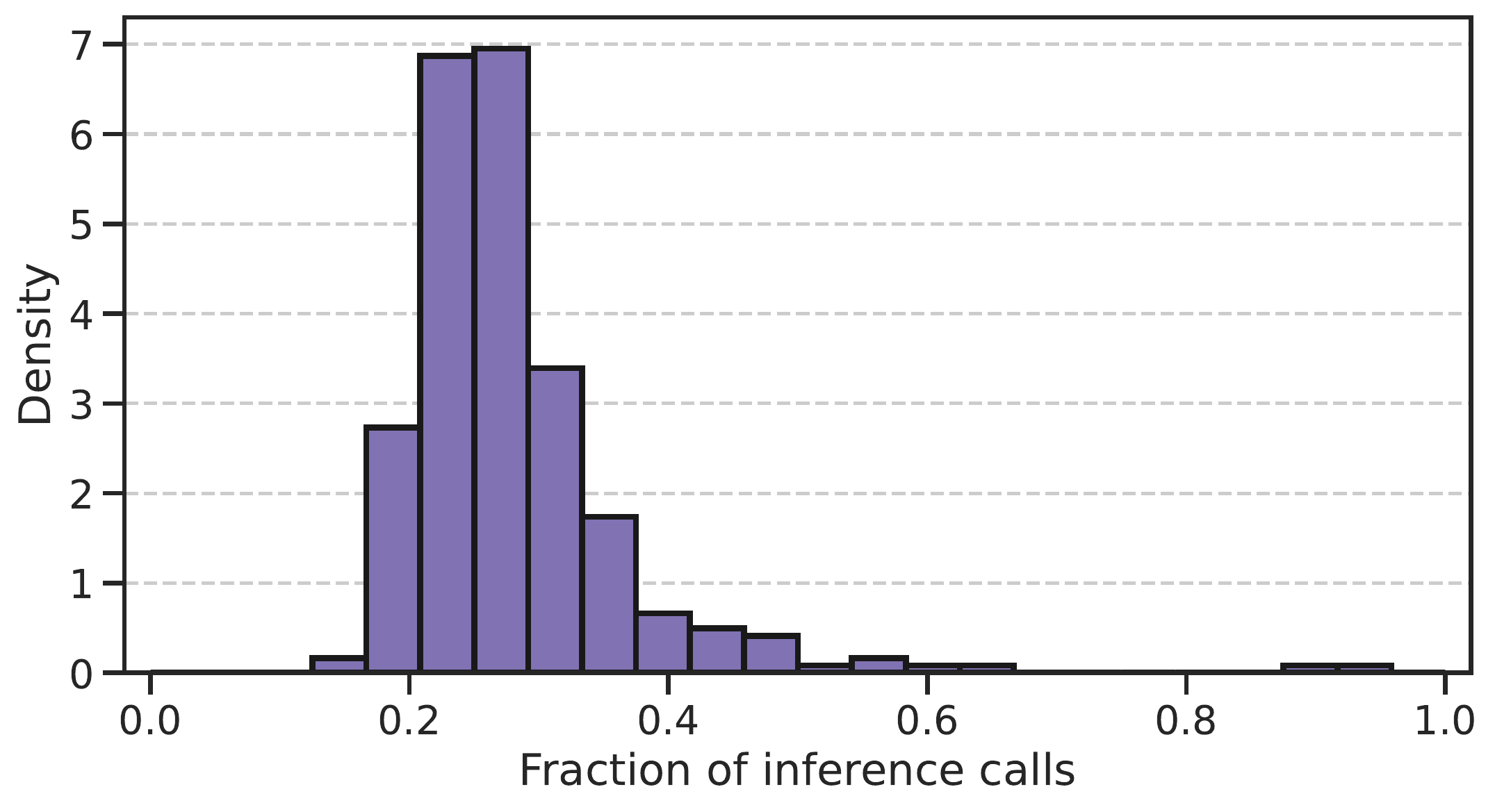}
    \caption{Histogram of the total number of inference calls made using DSP relative to the baseline optimization for a 0.2\% batch size over eight iterations.}
    \label{fig:dockstring-calls-002}
\end{figure}

\FloatBarrier

\begin{figure}
    \centering
    \includegraphics[height=0.8\textheight]{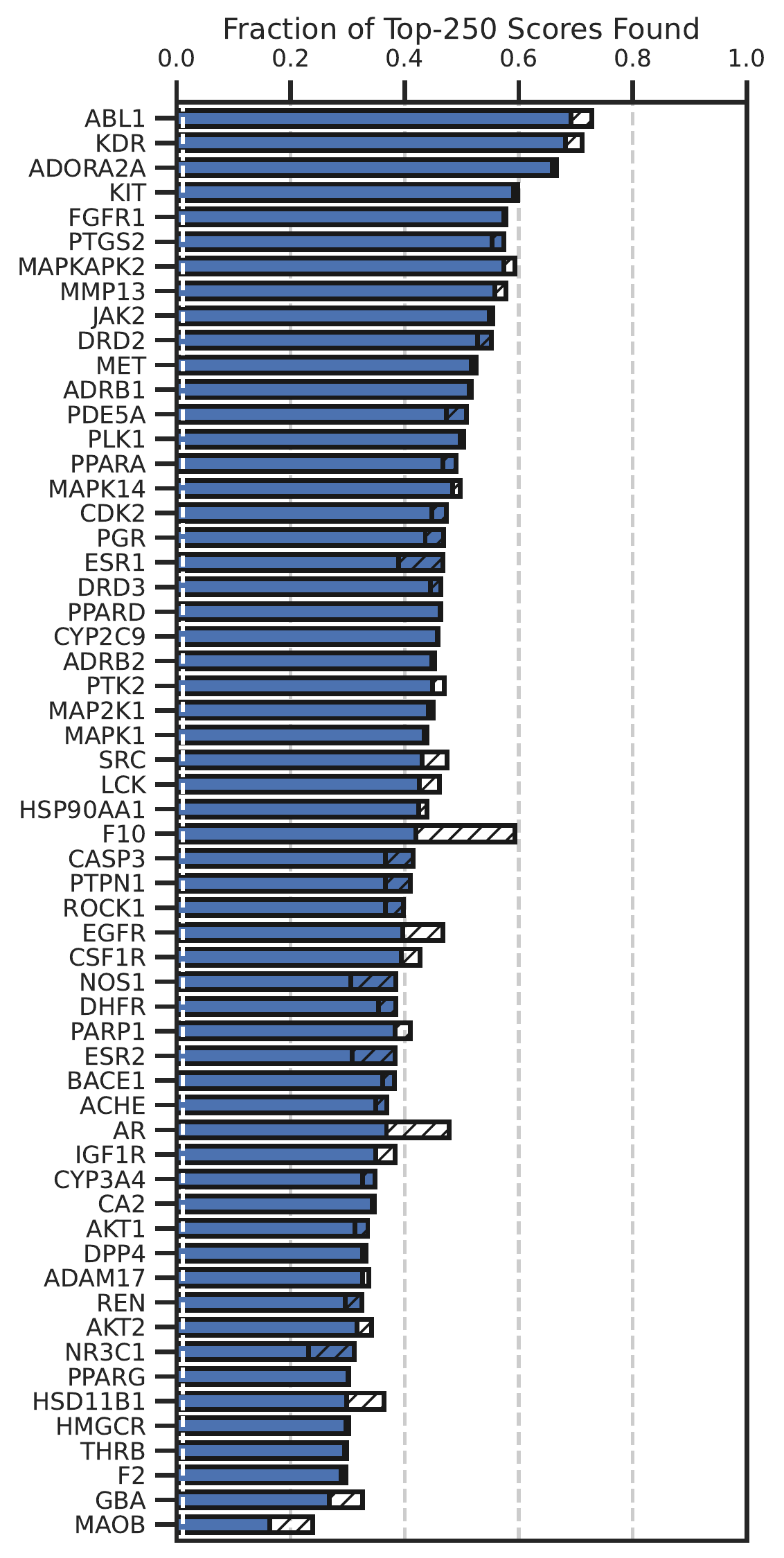}
    \caption{Bayesian optimization performance on DOCKSTRING tasks using DSP, an MPN surrogate model, UCB acquisition function, and 0.1\% batch size. Each bar indicates the average fraction of the top-250 compounds found during five independent runs with docking scores against given target protein as the objective function. Hatched areas indicate the difference in mean performance of the baseline optimization relative to DSP. The white, dotted vertical line represents expected performance of a random search. $\ast$ indicates targets where optimization without pruning outperforms DSP according to a one-sided $t$-test, $p\mathrm{-value}<0.05$ (Bonferroni corrected).}
    \label{fig:dockstring-bar-001}
\end{figure}

\begin{figure}
    \centering
    \includegraphics[height=0.8\textheight]{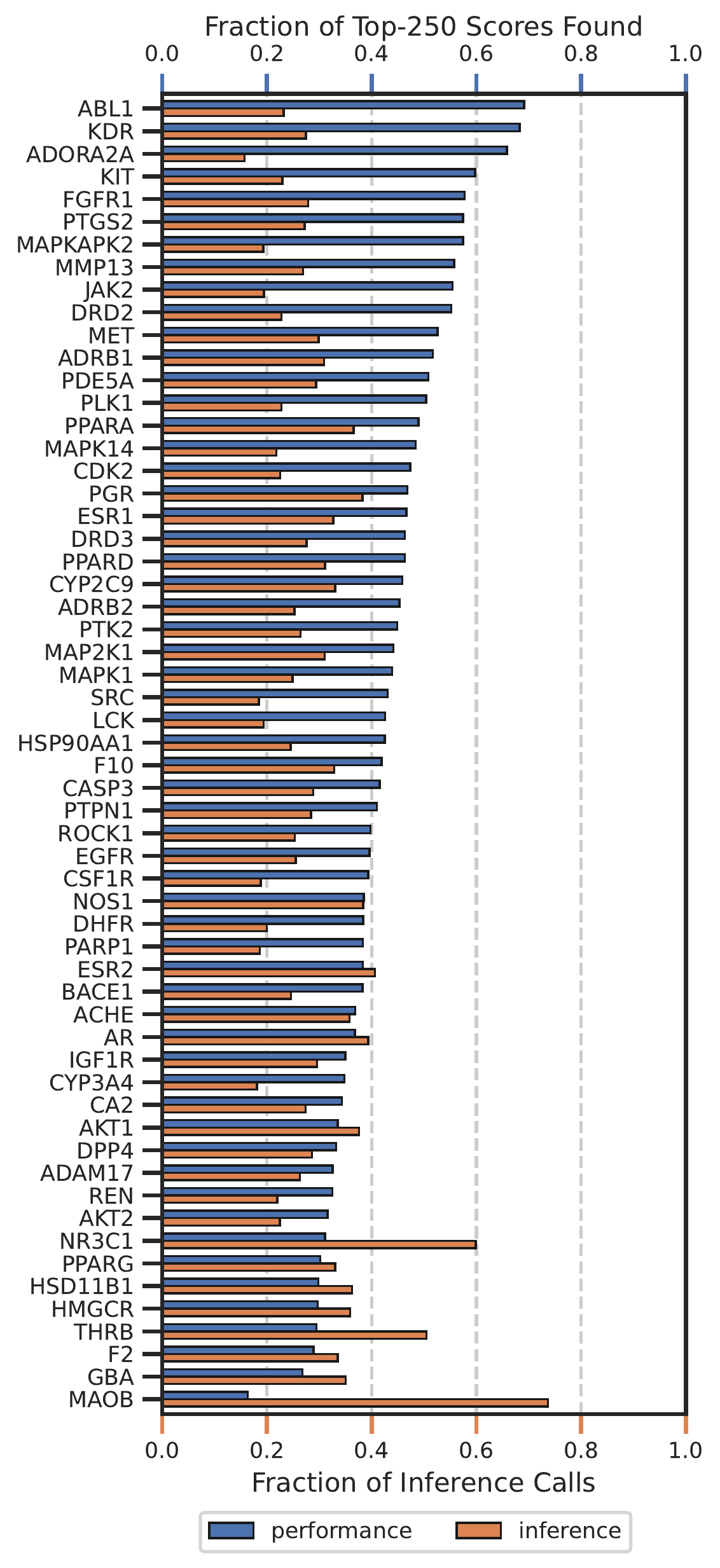}
    \caption{Bayesian optimization performance using DSP (top axis) and total number of inference calls using DSP relative to baseline optimization on DOCKSTRING tasks using a 0.1\% batch size over ten iterations. Each bar represents the mean of five independent runs.}
    \label{fig:dockstring-bar+calls-001}
\end{figure}

\begin{figure}
    \centering
    \includegraphics[width=0.6\textwidth]{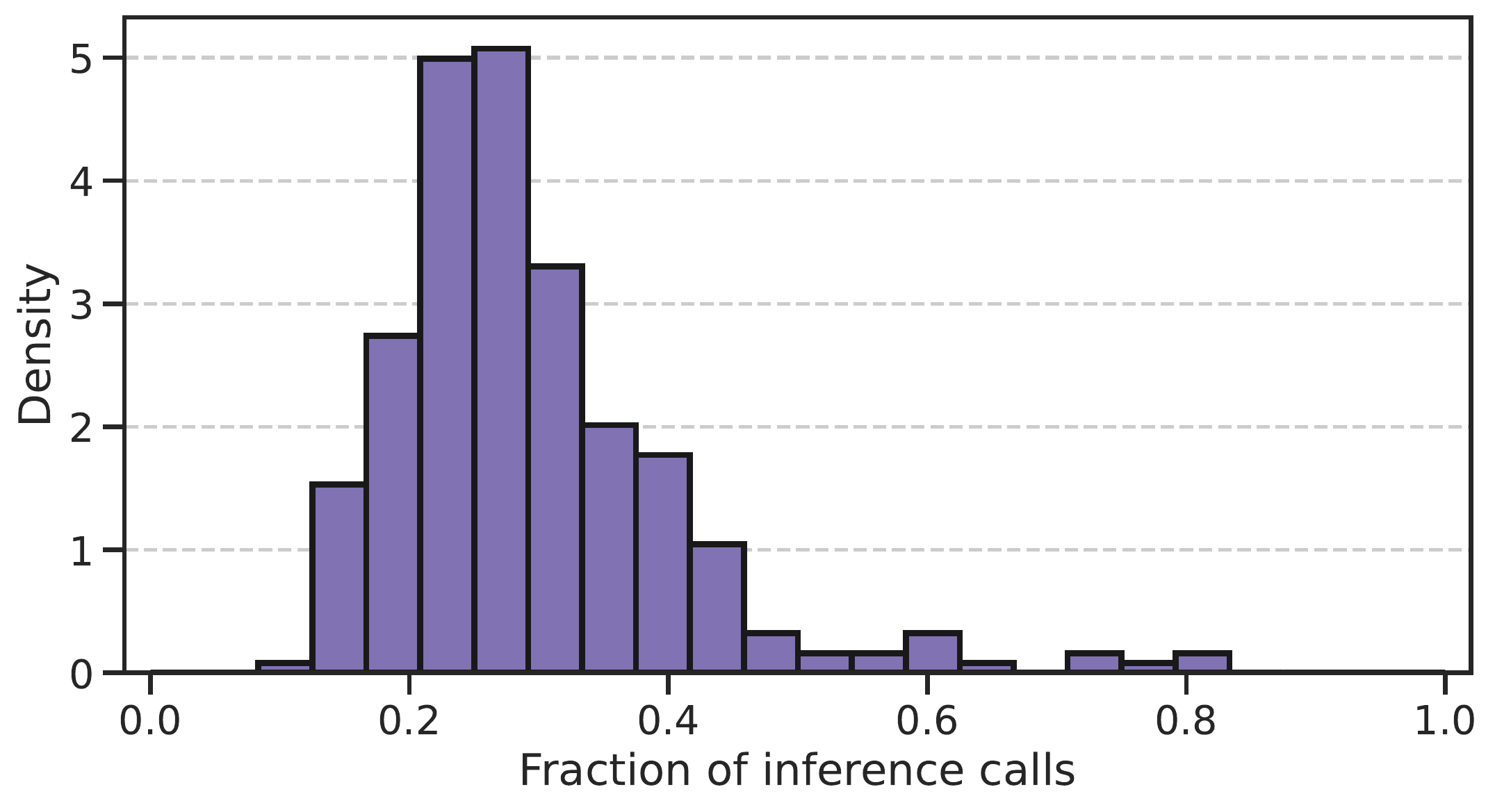}
    \caption{Histogram of the total number of inference calls made using DSP relative to the baseline optimization for a 0.1\% batch size over ten iterations.}
    \label{fig:dockstring-calls-001}
\end{figure}

\FloatBarrier

\begin{figure}
    \centering
    \includegraphics[width=0.6\textwidth]{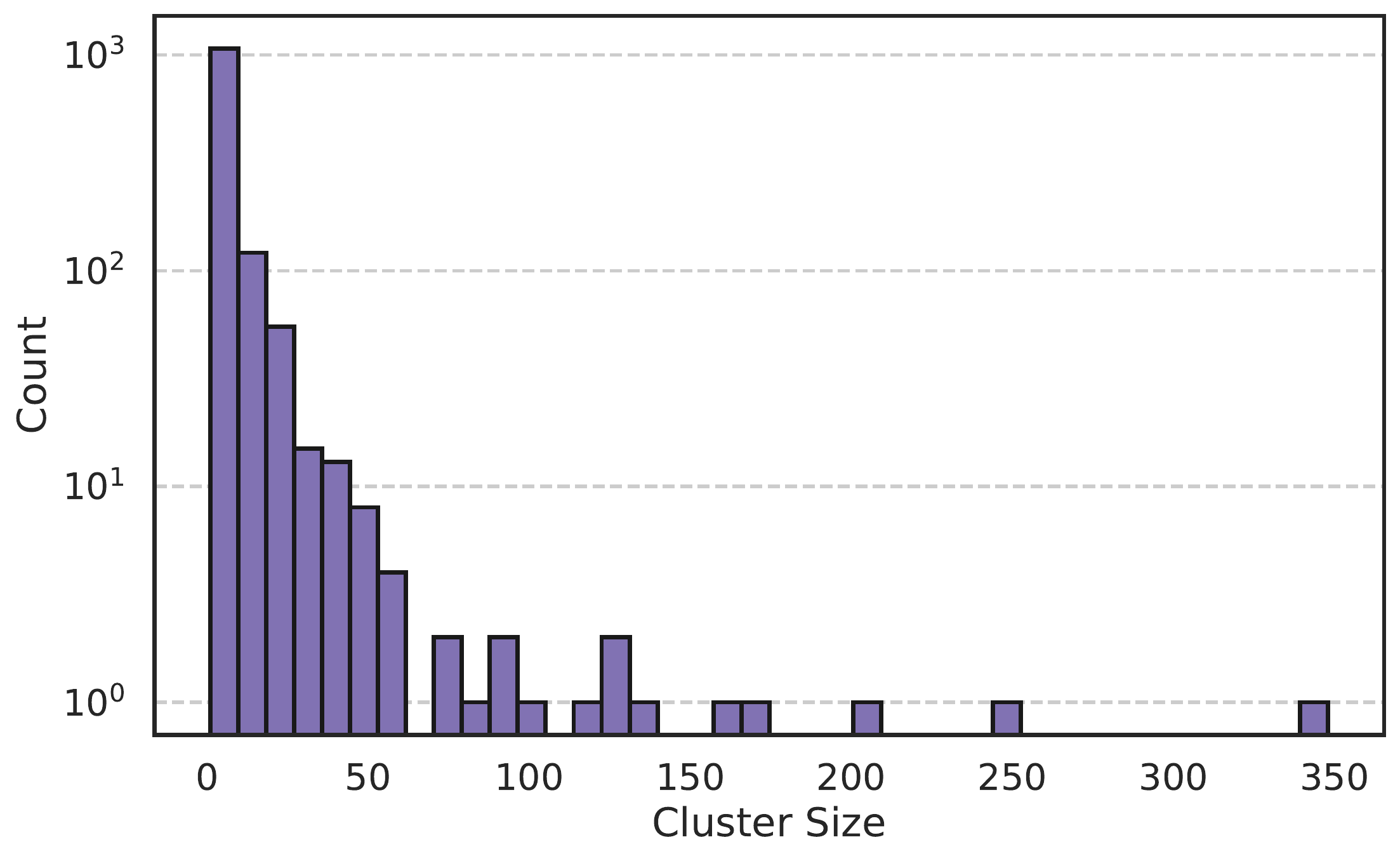}
    \caption{Histogram of cluster sizes in the top-\num{10000} molecules of the AmpC-Glide dataset using sphere clustering.} 
    \label{fig:ampc-clusters}
\end{figure}

\begin{figure}
    \centering
    \includegraphics[width=\textwidth]{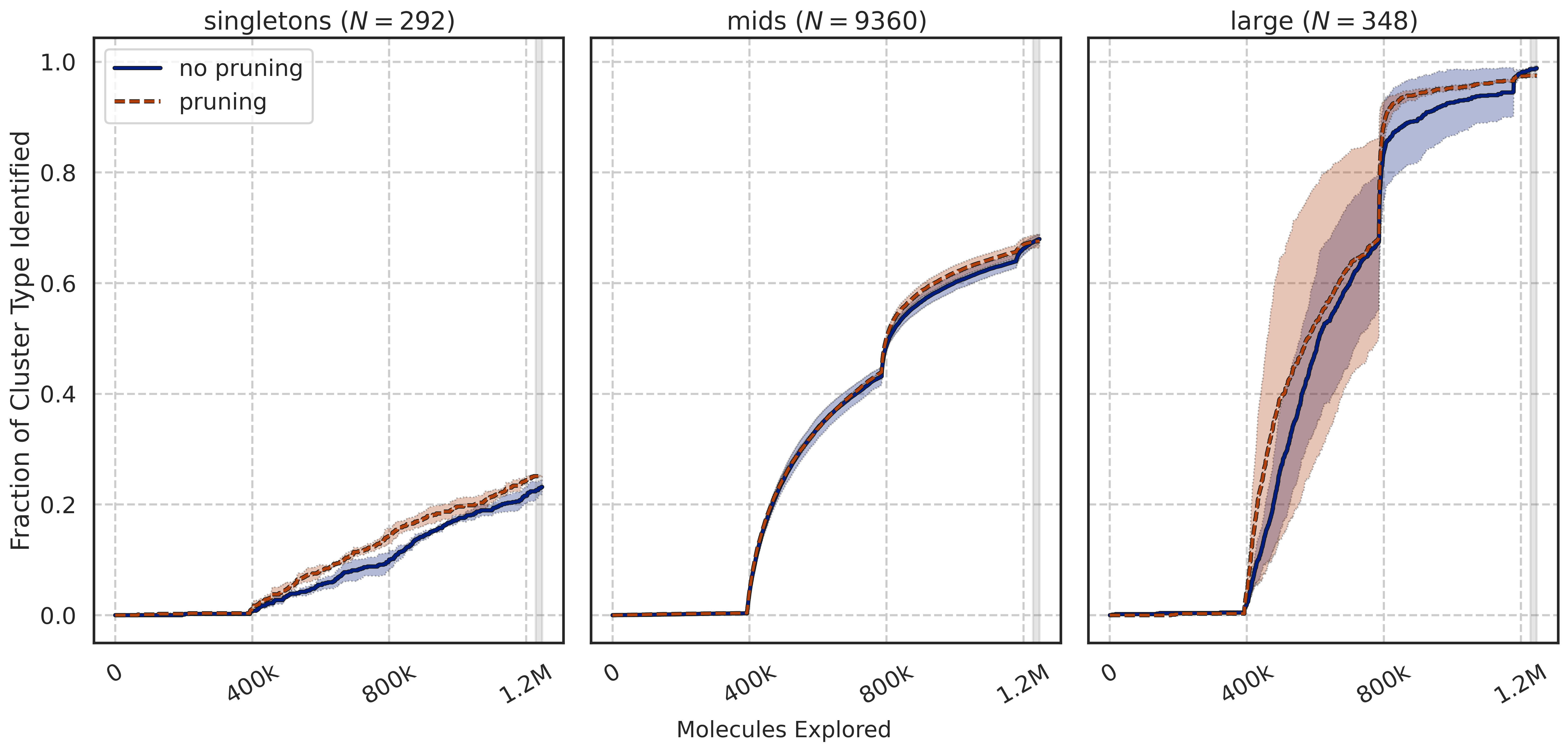}
    \caption{Fraction of molecules in each cluster type of the top-\num{10000} molecules identified versus number of molecules explored using a 0.4\% batch size. The grey shaded area represents the range of possible sample budgets for individual DSP trials. Each trace represents the average $\pm$ standard deviation of three independent runs.} 
    \label{fig:ampc-004-clusters}
\end{figure}

\FloatBarrier

\begin{figure}
    \centering
    \includegraphics[width=0.4\textwidth]{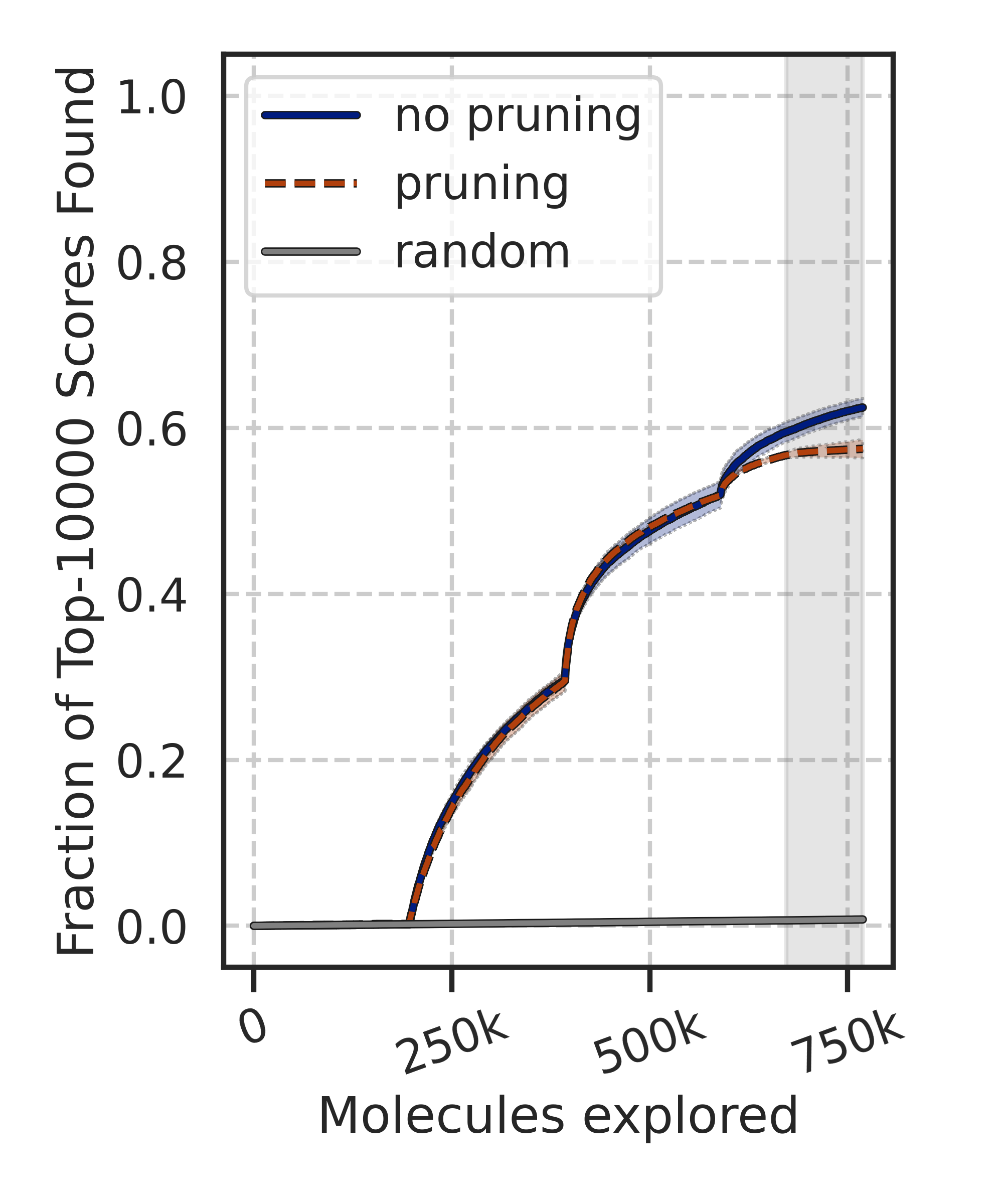}
    \caption{Bayesian Optimization performance on AmpC Glide dataset (100M) with an MPN surrogate model, UCB acquisition function, and 0.2\% batch size. The grey shaded area represents the range of possible sample budgets for individual DSP trials. Each trace represents the average $\pm$ standard deviations of three independent runs.} 
    \label{fig:ampc-002-perf}
\end{figure}

\begin{figure}
    \centering
    \includegraphics[width=\textwidth]{figures/si/ampc/clusters-002-10k.png}
    \caption{Fraction of molecules in each cluster type of the top-\num{10000} molecules identified versus number of molecules explored using a 0.2\% batch size. The grey shaded area represents the range of possible sample budgets for individual DSP trials. Each trace represents the average $\pm$ standard deviation of three independent runs.} 
    \label{fig:ampc-002-clusters}
\end{figure}

\FloatBarrier

\begin{figure}
    \centering
    \includegraphics[width=0.4\textwidth]{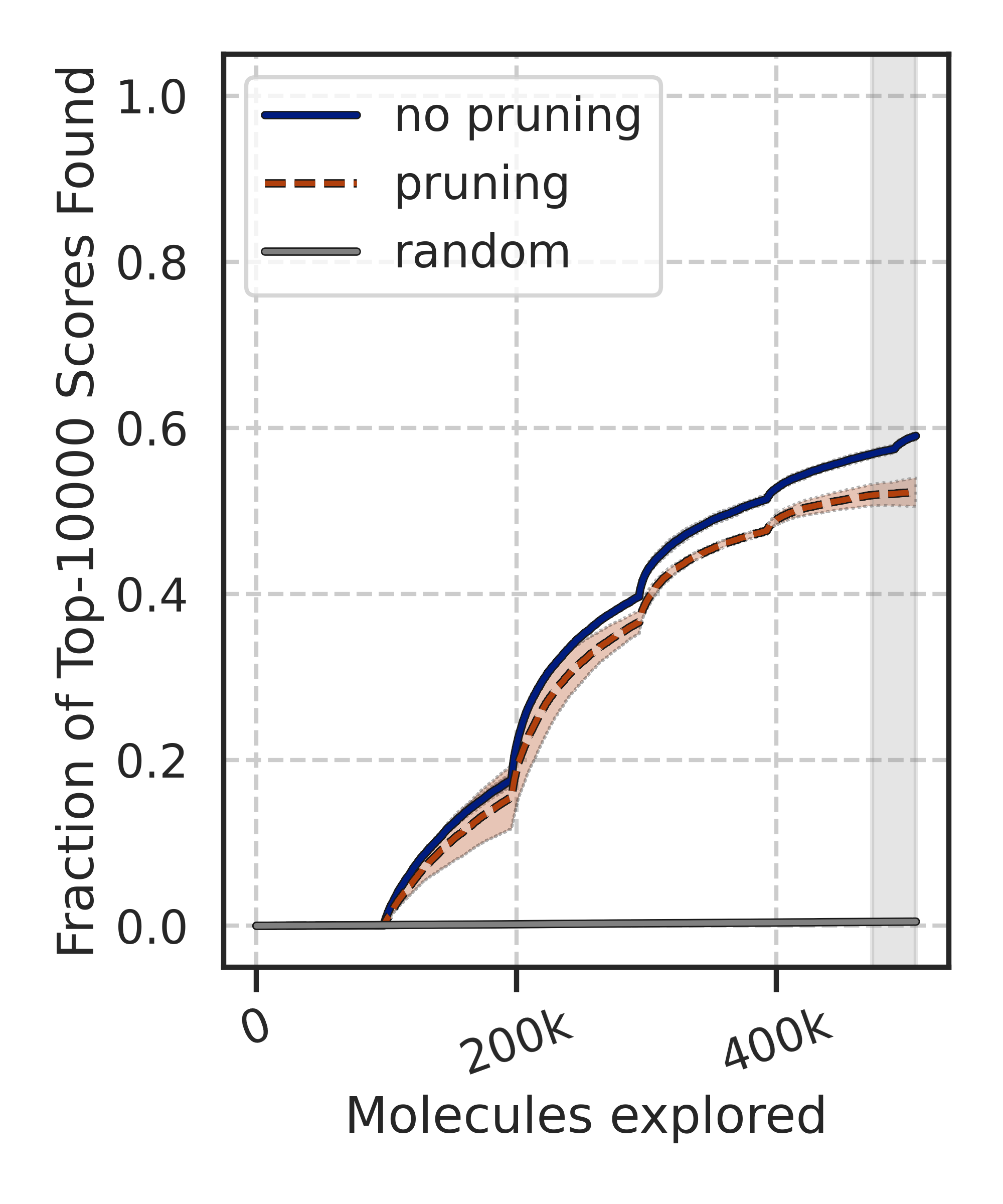}
    \caption{Bayesian Optimization performance on AmpC Glide dataset (100M) with an MPN surrogate model, UCB acquisition function, and 0.1\% batch size. The grey shaded area represents the range of possible sample budgets for individual DSP trials. Each trace represents the average $\pm$ standard deviations of three independent runs.} 
    \label{fig:ampc-001-perf}
\end{figure}

\begin{figure}
    \centering
    \includegraphics[width=\textwidth]{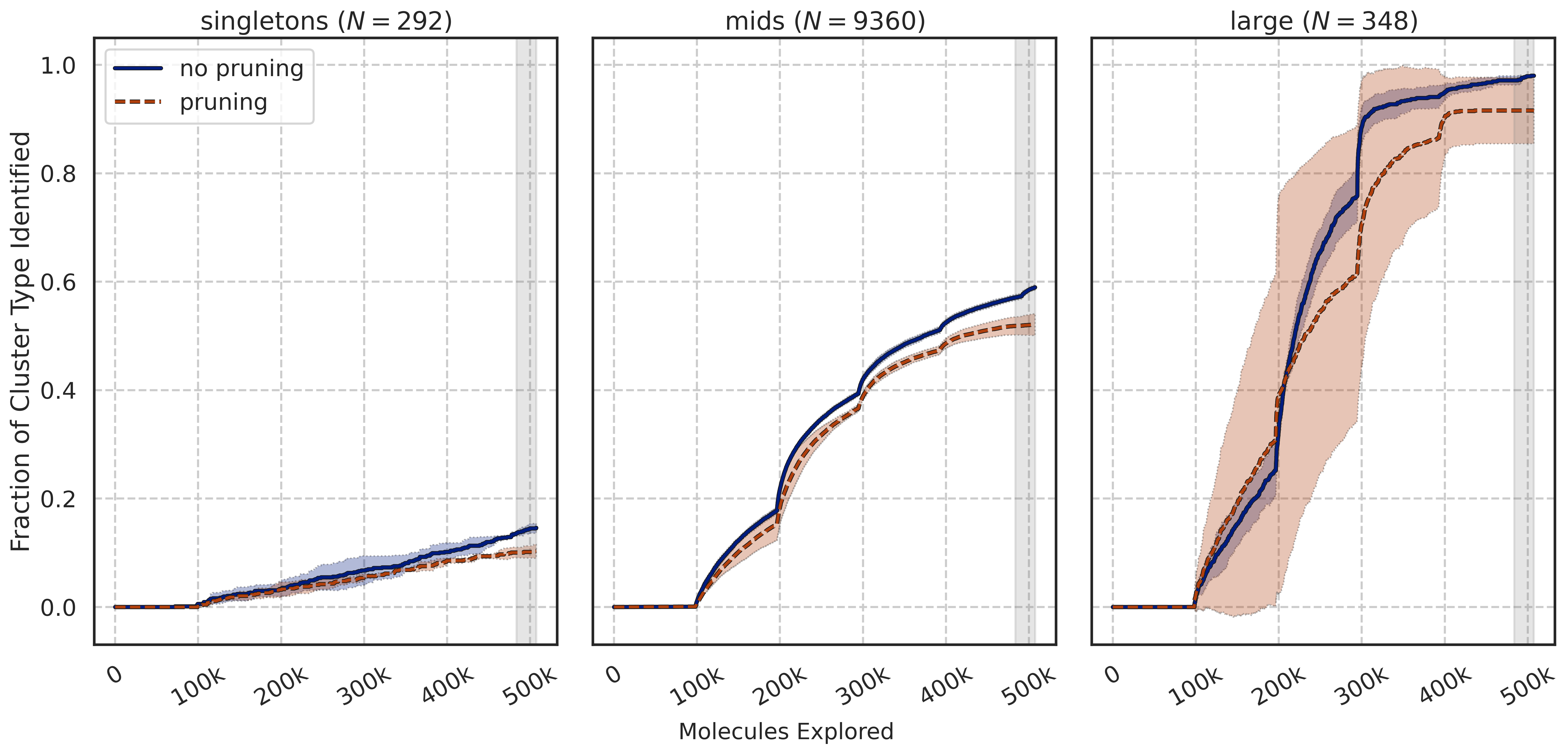}
    \caption{Fraction of molecules in each cluster type of the top-\num{10000} molecules identified versus number of molecules explored using a 0.1\% batch size. The grey shaded area represents the range of possible sample budgets for individual DSP trials. Each trace represents the average $\pm$ standard deviation of three independent runs.} 
    \label{fig:ampc-001-clusters}
\end{figure}